\documentclass{SciPost}

\hypersetup{
    colorlinks,
    linkcolor={red!50!black},
    citecolor={blue!50!black},
    urlcolor={blue!80!black}
}

\usepackage[bitstream-charter]{mathdesign}

\usepackage{amsmath}

\usepackage[scr=stixtwofancy]{mathalpha}

\newcommand{\bm}[1]{\boldsymbol{#1}}

\usepackage{cases}

\usepackage{tensor}

\usepackage{braket}

\usepackage{tcolorbox}
\usepackage{placeins}

\DeclareMathOperator{\Tr}{Tr}
\DeclareMathOperator{\diag}{diag}

\fancypagestyle{SPstyle}{
    \fancyhf{}
    \lhead{
        \colorbox{scipostblue}{
            \bfseries\color{white}~SciPost Physics~
        }
    }
    \rhead{
        \bfseries\color{scipostdeepblue}~Submission
    }
    
    \fancyfoot[C]{\textbf{\thepage}}
}

\begin{document}

\pagestyle{SPstyle}

\begin{center}{\Large \textbf{\color{scipostdeepblue}{
Hawking Radiation in non-Hermitian Microscopic Analogues\\
}}}\end{center}

\begin{center}\textbf{
Diego F. Munoz-Arboleda\textsuperscript{1$\star$},
Cristiane Morais Smith\textsuperscript{1} and
Marcus St\aa{}lhammar\textsuperscript{1,2}
}\end{center}

\begin{center}
{\bf 1} Institute for Theoretical Physics, Utrecht University, 3584CC Utrecht, The Netherlands
\\
{\bf 2} Department of Physics and Astronomy, Uppsala University, Uppsala, Sweden
\\[\baselineskip]
$\star$ \href{d.f.munozarboleda@uu.nl}{\small d.f.munozarboleda@uu.nl}
\end{center}

\section*{\color{scipostdeepblue}{Abstract}}
\textbf{\boldmath{%
Non-Hermitian systems have recently emerged as a convenient platform to realize black-hole physics. Here, we propose a microscopic open quantum system, namely a fermionic tight-binding chain coupled to Markovian reservoirs, which emulates black-hole properties. The resulting quadratic Lindbladian admits an effective non-Hermitian limit that produces a gain and loss lattice model with a non-reciprocal next-nearest-neighbor hopping. The rapidity spectrum of the full Lindbladian forms tilted exceptional cones that define an effective Painlevé-Gullstrand geometry and separates into black-hole and white-hole sectors, while steady-state particle densities and currents retain clear signatures at the horizon positions. We then analyze the effective non-Hermitian Hamiltonian that arise from the microscopic quantum system. The biorthogonal flux identifies the outgoing exterior and interior channels. Finally, we introduce a fermionic Gaussian Nambu extension and formulate Hawking-pair witnesses through frequency-resolved scattering, nonlocal density-density correlations, and anomalous Hawking-partner amplitudes. The resulting correlations display analogue Hawking radiation signatures, while the frequency-resolved Hawking-partner correlations satisfy the fermionic covariance-positivity constraint. These results establish a connection between microscopic open-system dynamics, emergent non-Hermitian geometry, and fermionic many-body probes of analogue Hawking radiation.
}}

\vspace{\baselineskip}

\noindent\textcolor{white!90!black}{%
\fbox{\parbox{0.975\linewidth}{%
\textcolor{white!40!black}{\begin{tabular}{lr}%
  \begin{minipage}{0.6\textwidth}%
    {\small Copyright attribution to authors. \newline
    This work is a submission to SciPost Physics. \newline
    License information to appear upon publication. \newline
    Publication information to appear upon publication.}
  \end{minipage} & \begin{minipage}{0.4\textwidth}
    {\small Received Date \newline Accepted Date \newline Published Date}%
  \end{minipage}
\end{tabular}}
}}
}


\vspace{10pt}
\noindent\rule{\textwidth}{1pt}
\tableofcontents
\noindent\rule{\textwidth}{1pt}
\vspace{10pt}


\section{Introduction}

Hawking radiation is a kinematic quantum-field effect associated with horizons: the strong redshift experienced by modes propagating near the horizon leads to a mixing between positive- and negative-frequency modes, and hence to pair creation \cite{hawking_particle_1975,Unruh1976,BirrelDavies-QFCST,wald1994QFTCST}. In the gravitational setting, this mechanism is most transparently understood by comparing the mode bases natural to different observers and coordinate systems. For a black hole formed by collapse, the physically relevant state is the Unruh vacuum, which is empty at past null infinity but regular on the future horizon, and therefore contains a net outgoing thermal flux at late times \cite{Unruh1976,Candelas1980,FabbriNavarro2005}. The emitted Hawking quantum is correlated with a partner mode that falls across the horizon. In this sense, Hawking radiation is not merely a one-body thermal spectrum but fundamentally a pair-production process.

The universality of this kinematic mechanism inspired the search for  systems that can mimic such gravitational signatures in the laboratory by realizing effective horizons in media whose collective excitations propagate as fields on curved spacetime backgrounds \cite{unruh_experimental_1981,barcelo_analogue_2011}. In acoustic and polariton fluid analogues, the relevant horizon is generated by the spatial variation of an effective flow velocity relative to the propagation speed of the elementary excitations \cite{unruh_experimental_1981,Visser1998,GarayEtAl2000,LahavEtAl2010,GeraceCarusotto2012,NguyenEtAl2015}. The analogue Hawking process then appears as a Bogoliubov mixing between positive- and negative-norm channels. For this reason, the most direct signatures of spontaneous analogue Hawking radiation are not only spectral occupations, but nonlocal correlations between the outgoing Hawking mode and its partner. In Bose-Einstein condensates and related bosonic platforms, these correlations appear as characteristic long-range structures in density-density correlators, sometimes described as Hawking ``moustaches'' \cite{BalbinotEtAl2008,CarusottoEtAl2008NJP,Steinhauer2015_PRD,deNova2019}.

For fermions, anomalous particle-partner correlations are constrained by positivity and Pauli exclusion. The corresponding normalized anomalous correlation must remain below unity as required by fermionic covariance positivity and Pauli exclusion.; satisfying this condition establishes that the covariance is compatible with a physical fermionic state rather than providing, by itself, a nonclassicality witness. Assessing the genuinely quantum character of such correlations requires an entanglement criterion adapted to the fermionic algebra \cite{ShapourianShiozakiRyu2017,ShapourianRyu2019}.

In recent years, there has been increasing interest in understanding the dynamics of horizons and Hawking radiation in systems surrounded by an environment, in both gravitational and analogue black-hole scenarios \cite{SinhaRavalHu2003,YuZhang2008,KaplanekBurgess2021,WusterSavage2007,LombardoTuriaci2012,LombardoTuriaci2013}. Open quantum systems provide a microscopic setting in which non-Hermitian effective models arise, thus requiring a description in terms of a biorthogonal formulation of quantum mechanics, with different left and right eigenvectors \cite{brody2014biorthogonal}. Markovian dynamics are described by the Franke-Gorini-Kossakowski-Sudarshan-Lindblad master equation \cite{Franke1976,GoriniKossakowskiSudarshan1976,Lindblad1976}, whose conditional quantum trajectories are governed between jumps by an effective non-Hermitian Hamiltonian \cite{DalibardCastinMolmer1992}. For quadratic open systems, the full Lindblad generator can be reduced to a non-Hermitian damping matrix, whose complex rapidities determine relaxation, stability, and steady-state correlations \cite{Prosen2008ThirdQuantization}. This connection has revealed dynamical manifestations of non-Hermitian band topology, including chiral damping, Liouvillian skin effects, and experimentally accessible signatures of non-Hermitian boundary modes that remain visible in unconditional open-system evolution \cite{SongYaoWang2019,HagaEtAl2021,YangJiangBergholtz2022,YangZelenayovaMoligniniBergholtz2025}. These developments are particularly relevant to the analogue-gravity setting.

Although effective non-Hermitian systems can generate exceptional cones and analogue horizons through gain, loss, and spatially varying non-reciprocal hopping \cite{Stalhammar-NJP2023,MunozArboleda2026ThermodynamicsAnalogue}, a Lindblad realization allows us to determine whether these geometric and spectral structures persist once quantum jumps and the complete long-time dynamics are included. In particular, the rapidity spectrum and steady-state correlations provide a microscopic means of testing the survival of the analogue black-hole and white-hole sectors and their associated horizon signatures beyond the effective non-Hermitian approximation.

The purpose of the present work is to realize this behavior within an open quantum system setting. Starting from a Hermitian tight-binding chain coupled to Markovian reservoirs, we derive a fermionic quadratic Lindbladian which reproduces an effective Painlevé-Gullstrand geometry and supports analogue horizons. The short-time effective non-Hermitian limit produces a gain and loss (GL) lattice model with non-reciprocal next-nearest-neighbor (NNN) hopping. The full Lindblad dynamics can then be solved exactly using the quadratic fermionic formalism, allowing us to examine which aspects of the resulting spectral structure persist in the steady state. We subsequently analyze the effective non-Hermitian Hamiltonian at the single-particle level. We couple three chains in series which are identical with the exception that the first and the third has a NNN hopping $\kappa_1$ and the second has $\kappa_2$. The chains are connected by a NNN hopping $\kappa_3$. Under periodic boundary conditions, region 2 represents the outer part of a black hole and regions 1 (3) the inner part. Its long-wavelength structure supports analogue horizons controlled by the kink profile of the non-reciprocal hopping $\kappa_3$. The asymptotic left and right Bloch eigenvectors define a biorthogonal flux that selects the outgoing exterior and interior channels. In this way, the single-particle description emerges as the effective analogue-gravity application of the microscopic open-system construction and fixes the analogue-horizon background and its kinematics.

While the microscopic open-system construction and its effective single-particle description identify the analogue horizon and its persistence in the steady state, they do not determine whether the outgoing channels exhibit Hawking pair correlations. We therefore introduce a fermionic Gaussian Nambu extension with an antisymmetric pairing matrix and formulate the stationary many-body problem as a frequency-resolved Bogoliubov scattering process. The resulting amplitudes determine the Hawking and partner occupations, their anomalous correlation, and the real-space covariance, whose cross-horizon ridges follow the expected equal-travel-time trajectory and are strongly suppressed in the no-crossing and uniform controls. These results allow us to identify signatures of analogue Hawking radiation.

The remainder of this paper is organized as follows. In Sec.~\ref{sec:OpenFermionicsystems}, we construct the microscopic fermionic Lindblad model, derive its effective non-Hermitian limit, analyze the rapidity spectrum, and study steady-state horizon signatures in particle densities and currents. In Sec.~\ref{sec:SPNHTBH}, we interpret the effective single-particle non-Hermitian Hamiltonian in terms of an analogue Painlevé-Gullstrand geometry and identify the outgoing channels using the biorthogonal flux. In Sec.~\ref{sec:many_body_witnesses}, we introduce the effective fermionic many-body extension and formulate frequency-resolved, density-density, anomalous-channel, and covariance-consistency diagnostics of Hawking pair correlations. We summarize and discuss our results in Sec.~\ref{sec:concandout}.

\section{Black-hole analogues in an open fermionic system}
\label{sec:OpenFermionicsystems}

We begin from a microscopic open-system description in which a
non-Hermitian structure emerges from a Hermitian
fermionic lattice coupled to Markovian reservoirs. This construction
serves two complementary purposes. First, it identifies the GL, and non-reciprocal hopping terms responsible for the analogue
geometry as consequences of engineered dissipation. Second, it makes
it possible to go beyond the conditional no-jump dynamics and study
the complete trace-preserving Lindblad evolution, including quantum
jumps, relaxation, and the nonequilibrium steady state. We show
that the effective non-Hermitian limit reproduces the GL-NNN structure
associated with analogue horizons (Sec.~\ref{subsec:microscopic_lindblad}), while the exact fermionic
Lindbladian retains corresponding black-hole and white-hole
sectors in its rapidity spectrum (Secs.~\ref{subsec:exactfermlindblad} and \ref{BOMSEC}). The resulting steady-state correlations and currents (Secs.~\ref{subsec:MSCM} and \ref{subsec:ESALH}) then provide a direct test of whether the horizon structure survives in the full long-time dissipative dynamics, rather than only in the effective short-time description.

\subsection{Lindblad construction and effective non-Hermitian limit}
\label{subsec:microscopic_lindblad}

The coherent part of the microscopic open system is a Hermitian tight-binding chain with $N$ unit cells and two fermionic modes, $a_j$ and $b_j$, per
cell,
\begin{equation}
    H_{\rm TB}
    =
    \tau\sum_{j=1}^{N}
    \left[
        b_j^\dagger(a_j+a_{j+1})
        +{\rm h.c.}
    \right],
    \label{eq:TBChain}
\end{equation}
where $\tau$ is the nearest-neighbor hopping and periodic boundary
conditions (PBC) are assumed unless stated otherwise.  The environment is
described by the four linear jump operators
\begin{align}
    L_j^{(1)}
    &=
    \sqrt{2(\kappa-\gamma)}\,b_j^\dagger,
    &
    L_j^{(2)}
    &=
    \sqrt{2(\kappa+\gamma)}\,a_j^\dagger,
    \nonumber\\
    L_j^{(3)}
    &=
    \sqrt{\kappa}\,
    \left(i b_j+b_{j+1}\right),
    &
    L_j^{(4)}
    &=
    \sqrt{\kappa}\,
    \left(a_j+i a_{j+1}\right),
    \label{eq:Dissipators}
\end{align}
with $\kappa,\gamma\in\mathbb R$ and $\kappa>|\gamma|$, so that all rates are real and non-negative. The two sublattices are distinguished by the different local and bond processes acting on them.

The reduced density matrix obeys the Lindblad equation
\begin{equation}
    \dot\rho = \mathcal L\rho = -i[H_{\rm TB},\rho] + \sum_{j,\mu} \left(
        L_j^{\mu}\rho L_j^{\mu\dagger}- \frac{1}{2} \left\{L_j^{\mu\dagger}L_j^{\mu},\rho \right\} \right).
    \label{eq:Lindblad}
\end{equation}
Along a conditional trajectory with no detected quantum jumps, the
unnormalized state evolves according to
\begin{align}
 \dot{\rho} &= -i[H_{\text{nH}}^{\text{eff}},\rho]=-i\left(H_{\text{nH}}^{\text{eff}}\rho-\rho H_{\text{nH}}^{\dagger\text{eff}}\right),
\end{align}
where $H_{\text{nH}}^{\text{eff}}$ is an effective nH Hamiltonian given by 
\begin{equation}
    H_{\rm nH}^{\rm eff}
    =
    H_{\rm TB}
    -
    \frac{i}{2}
    \sum_{j,\mu}
    L_j^{\mu\dagger}L_j^{\mu}.
    \label{eq:EffNHHam_def}
\end{equation}
Using the fermionic anticommutation relations, Eq.~\eqref{eq:EffNHHam_def}
becomes
\begin{align}
    H_{\rm nH}^{\rm eff}
    &=
    H_{\rm nH}[\kappa]
    -
    2i\kappa N,
    \label{eq:EffNHHam}
    \\
    H_{\rm nH}[\kappa]
    &=
    H_{\rm TB}
    +
    i\gamma
    \sum_{j=1}^{N}
    \left(
        a_j^\dagger a_j-b_j^\dagger b_j
    \right)
    +
    \frac{\kappa}{2}
    \sum_{j=1}^{N}
    \left(
        -a_{j+1}^\dagger a_j
        +a_j^\dagger a_{j+1}
        +b_{j+1}^\dagger b_j
        -b_j^\dagger b_{j+1}
    \right).
    \label{sp_hamiltonian}
\end{align}
The scalar term $-2i\kappa N$ shifts the energies to the lower-half complex plane, ensuring decaying solutions. The $\gamma$ and $\kappa$ parameters quantify the Markovian reservoir in the Lindbladian picture above, while, in the effective Hamiltonian they are related to the GL and non-reciprocal NNN structure, respectively.

Following the procedure in Ref.~\cite{MunozArboleda2026ThermodynamicsAnalogue}, an analogue-horizon background is engineered by implementing $\kappa\rightarrow f(r)$ in the lattice structure of the non-Hermitian effective Hamiltonian Eq.~\ref{sp_hamiltonian}, considering two chains, one with $\kappa_1$ and the other with $\kappa_2$ which are connected in a partition such that, we start with chain 1 ($\kappa_1$), then chain 2 ($\kappa_2$), and then back to chain 1 ($\kappa_1$) to be able to adopt PBC. The connections between chain 1 and chain 2 are given by $\kappa_3$. The profile that represents this setup is
\begin{equation}
    f(r) = \kappa_1 + \frac{\kappa_1-\kappa_2}{2} \left[\tanh\!\left(\frac{r_1-r}{l}\right) +\tanh\!\left(\frac{r-r_2}{l}\right)\right],
    \label{eq:horizonprofile}
\end{equation}
with $l$ the width of the domain wall. The profile approaches constant values in the asymptotic regions
\begin{numcases}{f(r)=}
    \sim \kappa_1 \quad &$r\ll r_1$, \nonumber \\
    \sim \kappa_2 \quad &$r_1\ll r \ll r_2$, \nonumber\\
    \sim \kappa_1 \quad &$r_2\ll r$
\end{numcases}
and contains interfaces at
\begin{equation}
f(r_1)=f(r_2)=\frac{\kappa_1+\kappa_2}{2}=\kappa_3.   
\end{equation}
This particular setup enables the possibility of building black-hole and white-hole horizon signatures in the surroundings of the interfaces $r_1$ and $r_2$, respectively.



\subsection{Exact fermionic Lindblad spectrum}
\label{subsec:exactfermlindblad}

We now turn to the unconditional long-time dynamics generated by
Eq.~\eqref{eq:Lindblad}.  In this description, $f(r)$ is a spatially
varying dissipative scale rather than a hopping amplitude imposed by
hand. The dynamics of a quadratic Lindbladian may be exactly solved using a technique known as {\it third quantization}~\cite{Prosen2008ThirdQuantization}. In the case of a fermionic system, the key step is to double the Hilbert space, and introduce a set of $4N$ Majorana fermions using the initial $2N$ complex fermions. We will outline the process here using the specific example given by Eqs.~\eqref{eq:TBChain} and \eqref{eq:Dissipators}, and refer to Refs.~\cite{SongYaoWang2019,HagaEtAl2021,YangJiangBergholtz2022,YangZelenayovaMoligniniBergholtz2025} for more general details.

From the initial fermionic basis $\{a_j,b_j\}_{j=1}^{N}$, we define a Majorana basis as
\begin{equation} \label{eq:Majoranabasis}
a_j =\frac{1}{2}\left(c_{j,a}-id_{j,a}\right) , \quad b_j = \frac{1}{2}\left(ic_{j,b}+d_{j,b}\right),
\end{equation}
where $c_{j,s}=c^{\dagger}_{j,s}$ and $d_{j,s}=d^{\dagger}_{j,s}$, with $s=\{a,b\}$. Collecting the Majorana basis in a vector, we find 
\begin{align}
    \bar{\omega}&=(\omega_{1},\omega_{2},...,\omega_{4N})^T =(c_{1,s}...,c_{N,s},d_{1,s},...,d_{N,s})^T.
\end{align}

Within this basis, the Hamiltonian of the Hermitian tight-binding chain takes the form
\begin{equation}
H_{\text{TB}} = \sum_{j,k}{\omega}^T_j \mathcal{H}_{j,k}{\omega}_k = \begin{pmatrix}\bar{c}^T&\bar{d}^{\,T} \end{pmatrix} \begin{pmatrix} \mathcal{H}_0 &0\\0&\mathcal{H}_0 \end{pmatrix} \begin{pmatrix}\bar{c}\\ \bar{d}\end{pmatrix}
\end{equation}
with $\bar{c}=(c_{1,s},...,c_{N,s})^T,\bar{d}=(d_{1,s},...,d_{N,s})^T$ and
\begin{equation}
\mathcal{H}_0 = \frac{i}{4} \begin{pmatrix} 0&\tau&0&0&\cdots &\tau\\-\tau& 0 &-\tau&0 & &0\\0&\tau&0&\tau & & \vdots\\ \vdots & &\ddots & \ddots& \ddots&0\\ 0&0 &\cdots &-\tau &0 &-\tau\\-\tau &0 &\cdots & 0&\tau&0 \end{pmatrix}.
\end{equation}
Similarly, the dissipators are decomposed as
\begin{equation}
L^{\mu}  = \sum_j \ell^{\mu}_j \omega_j,
\end{equation}
the components of which are collected to define a matrix $M$ (the use of which will become apparent shortly),
\begin{equation}
M = \sum_{\mu}M^{\mu}, \quad M_{\mu} = \left(\ell^{\mu}\right)^T \left(\ell^{\mu}\right)^*.
\end{equation}

In fermionic systems, the Lindbladian can be decomposed into an even and odd parity sector. Recalling that physical observables require an even number of fermionic operators, we may consider only the even parity sector of the fermionic parity operator $\mathcal{P}_F=(-1)^{\hat{\mathcal{N}}}=1$, with $\hat{\mathcal{N}}=\sum_j\varphi^\dagger_j\varphi_j=\bar{\varphi}^{\dagger}\cdot\bar{\varphi}$, the fermionic number operator \cite{YangJiangBergholtz2022}. Then, the physically relevant part of the Lindbladian can be put in the form
\begin{equation} \label{eq:EPLindblad}
\hat{\mathcal{L}}_+ = \frac{1}{2}\begin{pmatrix}\bar{\varphi}^{\dagger}& \bar{\varphi}\end{pmatrix} \begin{pmatrix} -X^{\dagger} & iY\\ 0 &X\end{pmatrix} \begin{pmatrix} \bar{\varphi} \\ \bar{\varphi}^{\dagger} \end{pmatrix} -A_0
\end{equation}
with $X = -4iH+M+M^T$, $Y = -2i\left(M-M^T\right)$ and $A_0 = \frac{1}{2}\Tr\left(X\right)$.
The adjoint basis of $\varphi$-fermions is defined as
\begin{align}
\varphi_j \ket{\omega_1^{\alpha_1}\cdots \omega_{2n}^{\alpha_{2n}}} &= \delta_{\alpha_j,1}\ket{\omega_j\omega_1^{\alpha_1}\cdots \omega_{2n}^{\alpha_{2n}}},
\\
\varphi_j^{\dagger} \ket{\omega_1^{\alpha_1}\cdots \omega_{2n}^{\alpha_{2n}}} &= \delta_{\alpha_j,0}\ket{\omega_j\omega_1^{\alpha_1}\cdots \omega_{2n}^{\alpha_{2n}}},
\\
\left\{\varphi_j,\varphi_k^{\dagger}\right\}  &= \delta_{j,k},
\end{align}
where $n=2N$ and $\alpha_j \in \{0,1\}$. The important observation from Eq.~\eqref{eq:EPLindblad}, is that the spectrum of the even parity-part of the Lindbladian coincides (up to a sign) with the spectrum of the damping matrix $X$, which in our specific case is defined by
\begin{equation}
 X =    -4i\mathcal{H} + \begin{pmatrix} M_1+M_1^T& -i\left(M_2-M_2^T\right)\\ i\left(M_2-M_2^T\right) & M_1+M_1^T \end{pmatrix},
\end{equation}
with constituents
\begin{equation}
    M_1 = f(r)\cdot \mathbb{I} +\frac{1}{2} \begin{pmatrix} \gamma&0&-i\frac{f(r)}{2} &0 &0&0&\cdots & i\frac{f(r)}{2}&0
    \\
    0 &-\gamma&0&i\frac{f(r)}{2} &0&0 &\cdots  &0&-i\frac{f(r)}{2}
    \\
    i\frac{f(r)}{2} & 0 &\gamma & 0&-i\frac{f(r)}{2}&0&\cdots&0&0
    \\
    0&-i\frac{f(r)}{2} & 0&-\gamma&0&i\frac{f(r)}{2}&\cdots&0&0
   \\
   \vdots & \ddots &\ddots&\ddots&\ddots&\ddots&\ddots&\vdots&\vdots
    \\
    0&0&\cdots &i\frac{f(r)}{2}&0&\gamma&0&-i\frac{f(r)}{2}&0
    \\
    0 & 0 & \cdots & 0&  -i\frac{f(r)}{2}&0&-\gamma&0&i\frac{f(r)}{2}
    \\
    -i\frac{f(r)}{2} & 0 & \cdots& 0 &0 & i\frac{f(r)}{2}&0&\gamma&0
    \\
    0&i\frac{f(r)}{2}&\cdots  &0 &0 &0 & -i\frac{f(r)}{2}&0&-\gamma\end{pmatrix},
\end{equation}
and $M_2 = M_1-f(r)\cdot\mathbb{I}$.
The sums and differences take somewhat simplified forms, since
\begin{align}
    M_1+M_1^T &= 2f(r)\cdot \mathbb{I}+\diag\left(\gamma,-\gamma,...,\gamma,-\gamma\right),
    \\
    i\left(M_2-M_2^T\right) &= \begin{pmatrix} 0 &0&\frac{f(r)}{2}\\0&0&0&-\frac{f(r)}{2}
    \\
    -\frac{f(r)}{2}&0&0&0&\ddots
    \\
    0&\frac{f(r)}{2} &0 &0
    \\
    &&\ddots\end{pmatrix}. 
\end{align}

Similarly, the matrix $Y$ takes the form
\begin{equation}
    Y = -2i \begin{pmatrix} M_1-M_1^T & -i\left(M_2+M_2^T\right)\\i\left(M_2+M_2^T\right) & M_1-M_1^T\end{pmatrix},
\end{equation}
with $M_1-M_1^T = M_2-M_2^T$ and $M_2+M_2^T = M_1+M_1^T-2f(r)\cdot \mathbb{I}$.

To find the spectrum, we first block diagonalize $X$,
\begin{align}
    \tilde{X} &= UXU^{\dagger} = \begin{pmatrix} \tilde{X}_c &0\\0&\tilde{X_d}\end{pmatrix}, \quad U = \frac{1}{\sqrt{2}}\begin{pmatrix} 1&i\\i&1\end{pmatrix}, \label{eq:DampTrans}
    \\
    \tilde{X}_c &= -4i\mathcal{H}_0+M_1+M_1^T+M_2-M_2^T,
    \\
    \tilde{X}_d &= -4i\mathcal{H}_0 + M_1+M_1^T-M_2+M_2^T.
\end{align}
Acting with a subsequent transformation\\ $V=\diag\left(i,1,i,1,...,i,1\right)$, we find
\begin{align}
    iV^{-1}\tilde{X}_cV &= 2if(r)\cdot\mathbb{I} + 2i N - H^{\text{eff}}_{\text{nH}}\left[f(r)\right],
    \\
    iV^{-1}\tilde{X_d}V &= 2i f(r)\cdot\mathbb{I} -2i N -  H^{\text{eff}}_{\text{nH}}\left[-f(r)\right],
\end{align}
which gives
\begin{align}
\beta^c_j &= 2\left[f(r)+N\right]+iE_j[f(r)], 
\\
 \beta^d_j &= 2\left[f(r)-N\right]+iE_j[-f(r)],
\end{align}
where $E_j$ denote the eigenvalues of the effective non-Hermitian Hamiltonian in Eq.~\eqref{eq:EffNHHam}, and the spectrum of the full Lindblad operator is recovered upon reversing the overall sign.

There is one specifically notable feature in the rapidity spectrum, namely the differences between the blocks given by $\tilde{X}_c$ and $\tilde{X}_d$.
These two blocks essentially give rise to the exact same spectrum, but with one crucial difference; the sign of the NNN hopping term is reversed.
Using the insights of Ref.~\cite{MunozArboleda2026ThermodynamicsAnalogue}, this induces a black-hole and white-hole symmetry; the spectrum of $\tilde{X}_c$ essentially coincides with that of the non-Hermitian Hamiltonian used there to mimic a black-hole horizon, while $\tilde{X}_d$ instead results in the horizon of an analogue white hole.
This emergent symmetry will be studied more closely in the next subsection.
\\
\subsection{Bloch operators and momentum space exceptional cones}
\label{BOMSEC}
The symmetry between black and white hole analogues is best studied in momentum space. Under PBC, the above lattice description in real space can be Fourier transformed to achieve a Bloch description. Following Ref.~\cite{YangZelenayovaMoligniniBergholtz2025}, the corresponding Bloch version of the Lindblad operator takes the form
\begin{align} \label{eq:Lopper}
\mathcal{L}_+(q) &= \frac{1}{2} \sum_{q} \begin{pmatrix} \bar{\varphi}^{\dagger}(q) & \bar{\varphi}(-q) \end{pmatrix}  \begin{pmatrix}-X^{\dagger}(q) & i Y(q) \\ 0 & X(q) \end{pmatrix} \begin{pmatrix} \bar{\varphi}(q)\\ \bar{\varphi}^{\dagger}(-q) \end{pmatrix} -A_0,
\end{align}
where

\begin{align}
X(q) &= 2 f(r) \mathbb{I} \nonumber
\\
&+ \begin{pmatrix} i\tau \left\{\sin(q) \sigma^x - \left[1+\cos(q)\right] \sigma^y\right\} + \gamma \sigma^z & -if(r) \sin(q) \sigma^z \\ if(r) \sin(q) \sigma^z & i\tau \left\{\sin(q) \sigma^x - \left[1+\cos(q)\right] \sigma^y\right\} + \gamma \sigma^z \end{pmatrix},
\\ 
Y(q) &= -2 \begin{pmatrix} i f(r) \sin(q) \sigma^z & \gamma \sigma^z \\ -\gamma \sigma^z& i f(r) \sin(q) \sigma^z \end{pmatrix}.
\end{align}

The spectrum of the damping matrix is again found by writing it in a block-diagonal form.
Acting with the transformation in Eq.~\eqref{eq:DampTrans}, we find
\begin{align}
\tilde{X}(q) &= UX(q)U^{\dagger} = 2 f(r)\mathbb{I}+ \begin{pmatrix} \tilde{X}_{\tilde{c}}(q) & 0 \\ 0 &\tilde{X}_{\tilde{d}}(q)\end{pmatrix},
\\
\tilde{X}_{\tilde{c}}(q) &= i\left\{\tau \sin(q)\sigma^x-\tau\left[1+\cos(q)\right]\sigma^y\right\} +\left[\gamma-f(r)\sin(q)\right]\sigma^z,
\\
\tilde{X}_{\tilde{d}}(q) &= i\left\{\tau \sin(q)\sigma^x-\tau\left[1+\cos(q)\right]\sigma^y\right\}+\left[\gamma+f(r)\sin(q)\right]\sigma^z,
\end{align}

where we use the subscripts $\tilde{c}$ and $\tilde{d}$ to emphasize the basis change governed by the transformation $U$.
The corresponding eigenvalues, or rapidities, read
\begin{align}
\beta_{c}(q) &= 2f(r) \pm \sqrt{\left[\gamma- f(r)\sin(q)\right]^2-\left[2+2\cos(q)\right]\tau^2},
\\
\beta_{d}(q) &= 2f(r) \pm \sqrt{\left[\gamma+ f(r)\sin(q)\right]^2-\left[2+2\cos(q)\right]\tau^2},
\end{align}
and we find exceptional points when
\begin{align}
\gamma &= f(r)\sin(q) \pm \sqrt{2\tau^2\left[1+\cos(q)\right]}, \label{eq:BHECper}
\\
\gamma &= -f(r)\sin(q) \pm \sqrt{2\tau^2\left[1+\cos(q)\right]}. \label{eq:WHECper}
\end{align}
Following Ref.~\cite{MunozArboleda2026ThermodynamicsAnalogue} and investigating the behavior close to $q=\pi$, the sets of these exceptional points form cones in $(\gamma,k)$-space coinciding exactly with those of the effective non-Hermitian Hamiltonian in Eq.~\eqref{eq:EffNHHam} [recalling $\kappa \rightarrow f(r)$],
\begin{equation} \label{eq:EClingamma}
\gamma = \mp f(r) k \pm \sqrt{\tau^2k^2},
\end{equation}
where the two sign choices are independent of each other. From now on, we consider $\tau=1$ without loss of generality.
These exceptional cones may be related to analogue spacetimes \cite{Stalhammar-NJP2023,MunozArboleda2026ThermodynamicsAnalogue}, in this specific case taking the form
\begin{equation}
    ds^2=\left[1-f^2(r)\right]dt^2\pm2f(r)dtdr+(dr)^2.
    \label{eq:PGmetric}
\end{equation}

Due to the explicit spatial dependence of the function $f(r)$ in Eq~\ref{eq:horizonprofile}, the commutation relation $[q,X(q)]=q\partial f(r)/\partial r$ is nonvanishing. According to Ref.~\cite{MunozArboleda2026ThermodynamicsAnalogue}, the commutator gives subleading corrections close to the horizon, which is true for these parameter regimes 
\begin{eqnarray}
	\displaystyle \left|\frac{\kappa_2-\kappa_1}{2l}q\right|\ll|\gamma|, \quad \left|\frac{\kappa_2-\kappa_1}{2l}\right|\ll|\tau|.
    \label{sler2}
\end{eqnarray}

For a sufficiently large chain ($r_1\ll r_2$), it is possible to define parameter ranges where the model mimics the interior, exterior, and horizon of a Schwarzschild-like black hole. This allows to further bound the system parameters as
\begin{equation}
    \frac{\kappa_1-\kappa_2}{l}=\frac{1}{r_1},\hspace{0.2cm} \kappa_1>1, \hspace{0.2cm}\kappa_2<1 \hspace{0.2cm}\text{and} \hspace{0.2cm} \kappa_3=1.
\end{equation}

Notably, although the corresponding effective non-Hermitian Hamiltonian hosts an exceptional cone reminiscent of a black-hole spacetime, the full Lindblad spectrum exhibits a pair of cones reminiscent of a black [$\beta_c(q)$] and a white hole [$\beta_d(q)$].

Finally, we solve Eq.~\eqref{eq:EClingamma} for $k$, and find four different solutions,
\begin{align}
k_{\text{BH}}^{\pm} &= \frac{\gamma f(r)}{1-f^2(r)} \pm \sqrt{\frac{\gamma^2}{\left[1-f^2(r)\right]^2}},
\\
k_{\text{WH}}^{\pm} &= -\frac{\gamma f(r)}{1-f^2(r)} \pm \sqrt{\frac{\gamma^2}{\left[1-f^2(r)\right]^2}}.
\end{align}
At asymptotic infinity [which in our case corresponds to when $f(r)\to \kappa_2$], their signs decide whether the momentum branch describes that of a particle or an antiparticle; positive sign gives the particle branch solution and negative sign gives the antiparticle branch solution~\cite{MunozArboleda2026ThermodynamicsAnalogue}. Thus, $k^+_{\text{BH}}$ and $k^+_{\text{WH}}$ correspond to particle channels for the black and white hole, respectively, while $k^-_{\text{BH}}$ and $k^-_{\text{WH}}$ are the corresponding antiparticle channels.

\subsection{Momentum space correlation matrices}
\label{subsec:MSCM}
The pair of exceptional cones reminiscent of black-hole and white-hole spacetimes emerging in the Lindblad spectrum may result in that the features predicted using the corresponding effective non-Hermitian Hamiltonian in Ref.~\cite{MunozArboleda2026ThermodynamicsAnalogue} may be at odds at larger time scales.
In what follows, we will demonstrate how the analogue horizons at $r=r_1$ and $r=r_2$ behave at late times, using single-particle correlation functions.
The momentum space correlation matrix $C(q)$ is defined from the Lyapunov equation~\cite{HorstmannCiracGiedke2013,ZhangBarthel2022Criticality},
\begin{equation} \label{eq:Lyaponov}
 X^{\dagger}(q) C(q) + C(q) X(q) = iY(q).
 \end{equation}
 
By multiplying Eq.~\eqref{eq:Lyaponov} with $U$ from the left and $U^{\dagger}$ from the right [as defined in Eq.~\eqref{eq:DampTrans}], we instead arrive at a block-diagonal form
\begin{equation}
\tilde{X}^{\dagger}(q) \tilde{C}(q) + \tilde{C}(q) \tilde{X}(q) = i\tilde{Y}(q)
\end{equation}
with $\tilde{C}(q) = UC(q)U^{\dagger}$ and $\tilde{Y}(q) = UY(q) U^{\dagger}$.
Solving this for $\tilde{C}(q)$ gives
\begin{align}
\tilde{C}(q) &= \begin{pmatrix} \tilde{C}_{\tilde{c}}(q) & 0 \\ 0 & \tilde{C}_{\tilde{d}}(q) \end{pmatrix},
\\
\tilde{C}_{\tilde{c}}(q) &= A_-\begin{pmatrix}\gamma-f(r)\sin(q)-2 f(r) & 1+e^{iq} \\1+e^{-iq} & \gamma-f(r)\sin(q)+2f(r) \end{pmatrix},
\\
\tilde{C}_{\tilde{d}}(q) &= A_+ \begin{pmatrix} \gamma+f(r)\sin(q)-2f(r)& 1+e^{iq}\\ 1+e^{-iq}&\gamma+f(r)\sin(q)+2f(r) \end{pmatrix},
\\
A_{\pm} &= \frac{2 \left[\gamma \pm f(r) \sin(q)\right]}{4\left[1+\cos(q) + 2 f^2(r)\right]-\left[\gamma\pm f(r) \sin(q)\right]^2}.
\end{align}
Importantly, $\tilde{C}(q)$ may {\it not} be interpreted as a Majorana pairing function, since the operators
\begin{align}
\tilde{c}_{s}(q) &=  \frac{1}{\sqrt{2}} \left[ c_{s}(q) + i d_{s}(q)\right],
\\
\tilde{d}_{s}(q) &= \frac{1}{\sqrt{2}} \left[ic_{s}(q)+d_{s}(q)\right],
\end{align}
are not Majorana operators (recall that $s=\{a,b\}$); $\tilde{c}_{s}(q) = \pm i\tilde{d}^{\dagger}_{s}(q)\neq \tilde{c}^{\dagger}_{s}(q)$.
To recover this interpretation, we transform back to the Majorana basis and find
\begin{equation} \label{eq:SScovmat}
C(q) = \frac{1}{2}\begin{pmatrix} \tilde{C}_{\tilde{c}}(q) + \tilde{C}_{\tilde{d}}(q) & i\left[\tilde{C}_{\tilde{c}}(q)-\tilde{C}_{\tilde{d}}(q)\right]\\-i\left[\tilde{C}_{\tilde{c}}(q)-\tilde{C}_{\tilde{d}}(q)\right] & \tilde{C}_{\tilde{c}}(q) + \tilde{C}_{\tilde{d}}(q) \end{pmatrix}.
\end{equation}
The elements of $C(q)$ are related to the Majorana-steady-state expectation values as~\cite{YangJiangBergholtz2022},
\begin{equation}
C_{lm}(q) = -\langle \omega_l(q)\omega_m(q) \rangle + \delta_{lm}
\end{equation}
where $\bar{\omega}(q):=\left[c_s(q),d_s(q)\right]$.
The physical picture is recovered through the fermionic correlation matrix.
Recalling Eq.~\eqref{eq:Majoranabasis}, the fermionic steady-state pairing functions can be collected in a correlation matrix $Q$,
\begin{align} \label{eq:fermcorr}
Q &= \begin{pmatrix} \langle a^{\dagger}a\rangle_{\text{SS}}& \langle a^{\dagger}b\rangle_{\text{SS}}\\ \langle b^{\dagger}a\rangle_{\text{SS}}& \langle b^{\dagger}b\rangle_{\text{SS}} \end{pmatrix} -\frac{1}{4} \begin{pmatrix}  \left[ \tilde{C}_{\tilde{c}}^{11}-\tilde{C}_{\tilde{d}}^{11}\right] & -2i\tilde{C}_{\tilde{c}}^{12} \\ 2i \tilde{C}_{\tilde{c}}^{12} & \left[ \tilde{C}_{\tilde{c}}^{22}-\tilde{C}_{\tilde{d}}^{22} \right]\end{pmatrix} +\frac{1}{2} \mathbb{I},
\end{align}
where the $q$-dependence has been omitted for brevity.
\begin{figure}
\centering
    \includegraphics[width=\textwidth]{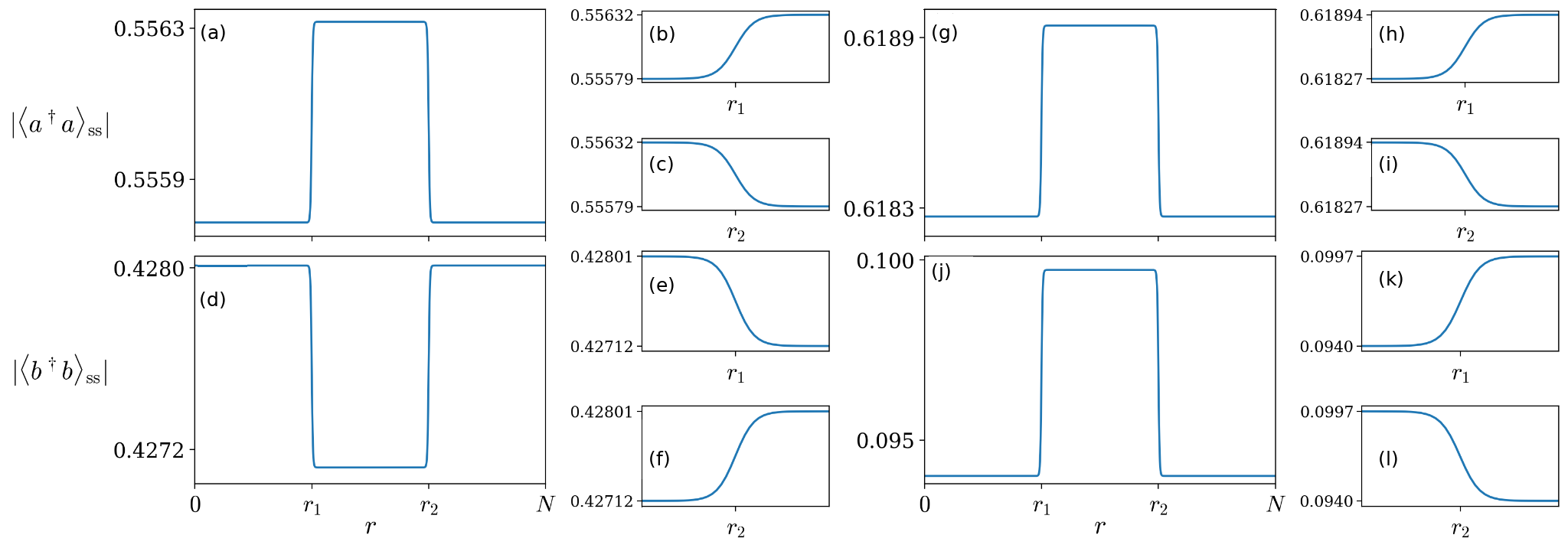}  
    \caption{Absolute values of the expected steady-state particle number densities of the open tight-binding chain described by Eq.~\eqref{eq:Lopper}; (a)-(c), (g)-(i) for $a$-fermions, and (d)-(f), (j)-(l) for $b$-fermions. All panels are restricted to momentum satisfying Eq.~\eqref{eq:BHECper}; the positive solution corresponding to the particle channel in (a)-(f), and the negative solution corresponding to the antiparticle channel in (g)-(l). There are sharp transitions at $r=r_1$ and $r=r_2$, indicating the existence of analogue black-hole event horizons in the steady state. For all panels, $r_1=100$, $r_2=200$, $N=300$, $l=1$, $\gamma=0.3$, and the zoomed-in plots (b), (e), (h), (k) are focused on the symmetric intervals $r\in \{r_1-5,r_1+5\}$ and (c), (f), (i), (l) in $r\in \{r_2-5,r_2+5\}$.} \label{fig:BH_Number}
\end{figure}
\begin{figure}
\centering
    \includegraphics[width=\textwidth]{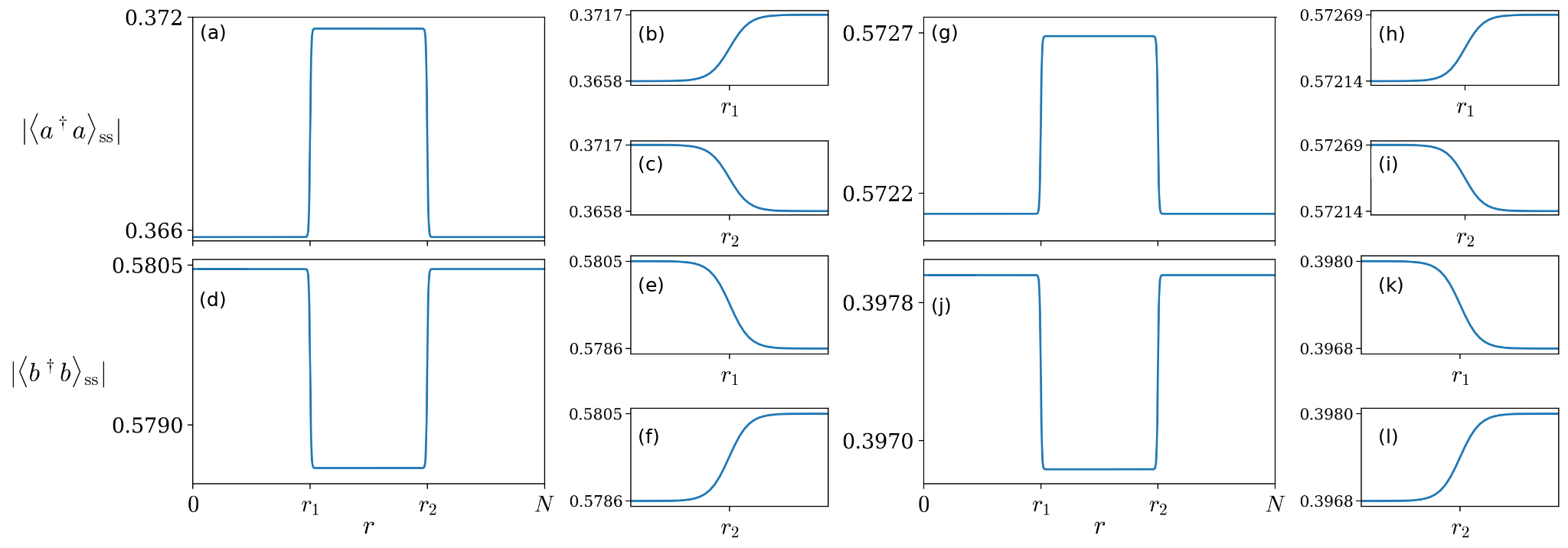}  
    \caption{Absolute values of the expected steady-state particle number densities of the open tight-binding chain described by Eq.~\eqref{eq:Lopper}; (a)-(c), (g)-(i) for $a$-fermions, and (d)-(f), (j)-(l) for $b$-fermions. All panels are restricted to momentum satisfying Eq.~\eqref{eq:WHECper}; the positive solution corresponding to the particle channel in (a)-(f), and the negative solution corresponding to the antiparticle channel in (g)-(l). There are sharp transitions at $r=r_1$ and $r=r_2$, indicating the existence of analogue white-hole event horizons in the steady state. For all panels, $r_1=100$, $r_2=200$, $N=300$, $l=1$, $\gamma=0.3$, and the zoomed-in plots (b), (e), (h), (k) are focused on the symmetric intervals $r\in \{r_1-5,r_1+5\}$ and (c), (f), (i), (l) in $r\in \{r_2-5,r_2+5\}$.} \label{fig:WH_Number}
\end{figure}

\subsection{Experimental signatures of the analogue Lindbladian horizon}
\label{subsec:ESALH}

We now turn to identifying experimental signatures indicating the existence of an analogue event horizon in the system described by Eq.~\eqref{eq:Lopper}. The purpose of this is not to derive additional features analogue to Hawking radiation, but rather to display the persistence of the horizon structure at large times. We will therefore focus on observables governed by the fermionic steady-state correlation matrix $Q$, treating equal-momentum single-particle correlations in Sec.~\ref{sec:momcorr}, and steady-state currents in Sec.~\ref{sec:current}.
\subsubsection{Equal-momentum correlations at exceptional cones}
\label{sec:momcorr}
The fermionic correlation matrix $Q$ provides us with momentum space versions of the steady-state number densities; $|\langle a^{\dagger}a\rangle_{\text{SS}}|= Q_{(1,1)}$ and $|\langle b^{\dagger}b\rangle_{\text{SS}}|= Q_{(2,2)}$.
Of particular interest in terms of the gravitational analogue that we are considering in this work, is exactly such equal-momentum correlations evaluated at momentum corresponding to the exceptional cones.
Figure.~\ref{fig:BH_Number} shows such number densities in the steady-state for momentum corresponding to the black-hole exceptional cone, while Fig.~\ref{fig:WH_Number} displays the case for momentum corresponding to the white-hole exceptional cone.
Both figures display a clear transition, in both $a$ and $b$ fermion number densities, at both the horizons.
Therefore, we can conclude that the horizons indeed persist individually in the steady state.
We note though that this does not provide information regarding whether or not these individual contributions cancel each other; this question is instead answered below.

\subsubsection{Fermionic steady-state current} 
\label{sec:current}

\begin{figure}[t]
\centering
    \includegraphics[width=\columnwidth]{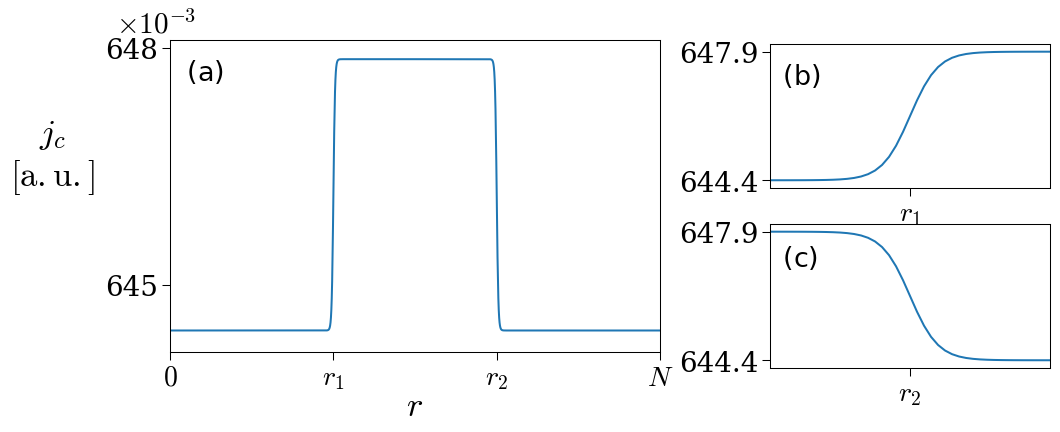}  
    \caption{Steady-state current given by Eq.~\eqref{eq:currentflow}.
    Panel (a) shows sharp variations at the two interfaces, $r=r_1$ and $r=r_2$, while panels (b) and (c) show enlarged views around the first and second interfaces, respectively. The persistence of these features after summing over momentum provides a steady-state signature of the analogue horizons.
    The parameters are $r_1=100$, $r_2=200$, $N=300$, $l=1$, and
    $\gamma=0.3$.} \label{fig:Current}
\end{figure}

Although the horizon structures become apparent studying the equal-momentum correlations discussed above, they do not comprise observables that are conventionally desired, especially considering the spatially varying momentum.
Using the fermionic correlation matrix, we may however also derive an analytical expression for the steady-state current flow, which does serve as an ideal observable in Lindblad systems~\cite{YangZelenayovaMoligniniBergholtz2025}.
Within the basis of the $\tilde{c}$ and $\tilde{d}$-operators, the current flow is defined as
\begin{align}
j_c &= j_c^{\tilde{c}}+j_c^{\tilde{d}},
\\
j_c^{\tilde{c}} &= \frac{i}{2N} \sum_{j=1}^N\left[ \langle \left(\tilde{c}^a_j\right)^{\dagger}\tilde{c}^b_{j}\rangle-\langle \left(\tilde{c}^b_j\right)^{\dagger}\tilde{c}^a_{j}\rangle+\langle \left(\tilde{c}^b_j\right)^{\dagger}\tilde{c}^a_{j+1}\rangle-\langle \left(\tilde{c}^a_{j+1}\right)^{\dagger}\tilde{c}^b_{j}\rangle\right],
\\
j_c^{\tilde{d}} &= \frac{i}{2N} \sum_{j=1}^N\left[ \langle \left(\tilde{d}^a_j\right)^{\dagger}\tilde{d}^b_{j}\rangle-\langle \left(\tilde{d}^b_j\right)^{\dagger}\tilde{d}^a_{j}\rangle+\langle \left(\tilde{d}^b_j\right)^{\dagger}\tilde{d}^a_{j+1}\rangle-\langle \left(\tilde{d}^a_{j+1}\right)^{\dagger}\tilde{d}^b_{j}\rangle\right].
\end{align}
Transforming back to $a$ and $b$-operators, and Fourier transforming them, after a straight-forward calculation we find,
\begin{equation}
j_c = \frac{2}{N} \sum_q \left[ \cos(q)-1\right] \text{Re}\left[ \langle b^{\dagger}(q)a(q)\rangle \right],
\end{equation}
where $\langle b^{\dagger}(q)a(q)\rangle$ may be read-off directly from the fermionic correlation matrix $Q$ in Eq.~\eqref{eq:fermcorr}.
For a sufficiently long chain, the corresponding sum (which takes the form of a Riemann sum) is approximated by an integral, and we conclude
\begin{equation} \label{eq:currentflow}
j_c \xrightarrow[N\to \infty]{} 2\int_{-\pi}^{\pi} dq \left[ \cos(q)-1\right] \text{Re}\left[ \langle b^{\dagger}(q)a(q)\rangle \right].
\end{equation}
The steady-state current is depicted in Fig.~\ref{fig:Current}, displaying a non-uniform behavior across the chain with a sharp transition happening exactly at the location of the horizons, which are zoomed in on in Fig.~\ref{fig:Current} (b) for $r=r_1$, and in Fig.~\ref{fig:Current} (c) for $r=r_2$.

We emphasize that we do not interpret this behavior as some kind of anomalous current flow or radiation, but rather as a clear experimental indication of the existence of the horizon structure; also at later times, the analogue horizons at $r=r_1$ and $r=r_2$ do not vanish when the system is allowed to evolve in time, indicating that the analogy to curved spacetime persists.
More importantly, even when summing (or integrating) over all momenta branches, the horizon remains; neither superposing the black-hole and white-hole branches nor summing the remaining momenta branches makes the horizon go away, making it a highly stable feature within our system.

\section{Single-particle analogue-gravity interpretation}
\label{sec:SPNHTBH}

Section~\ref{sec:OpenFermionicsystems} established that the conditional no-jump dynamics of the microscopic fermionic Lindbladian is generated, up to the scalar shift in Eq.~\eqref{eq:EffNHHam}, by the non-Hermitian operator $H_{\rm nH}[f]$ in Eq.~\eqref{sp_hamiltonian}. It showed how an effective geometry arise as a consequence of the coupling between the tight-binding chain with the Markovian reservoir in the rapidities spectrum, and that the associated interfaces remain visible in the steady state. To identify Hawking-like radiation, first we need to characterize the modes related to the Hawking and partner particles. This section uses the effective non-Hermitian Hamiltonian description to identify which asymptotic modes should be used to construct Hawking-partner observables.   


Following the same procedure and condition of Sec.~\ref{BOMSEC}, 
the Bloch Hamiltonian in the basis $(a_q,b_q)^{T}$ is
\begin{equation}
    \mathcal H_{\rm B}(q) = \tau[1+\cos(q)]\sigma^x + \tau\sin(q)\sigma^y +
    i[\gamma-f(r)\sin(q)]\sigma^z.
    \label{eq:single_particle_Bloch_exact}
\end{equation}
Its spectrum is
\begin{equation}
    E_\pm(q)
    =
    \pm
    \sqrt{
        2\tau^2[1+\cos(q)]
        -
        [\gamma-f(r)\sin(q)]^2
    }.
    \label{eq:single_particle_spectrum_exact}
\end{equation}
Expanding around $q=\pi$, with $k=q-\pi$, gives
\begin{equation}
    \mathcal H_{\rm B}(k)
    =
    -\tau k\,\sigma^y
    +
    i[\gamma+f(r)k]\sigma^z
    +
    \mathcal O(k^2),
    \label{eq:single_particle_Bloch_linear}
\end{equation}
and
\begin{equation}
    E_\pm(k)
    =
    \pm
    \sqrt{
        \tau^2k^2-[\gamma+fk]^2
    },
    \label{eq:single_particle_spectrum_linear}
\end{equation}
which lead us to the same exceptional cone structure as Eq.~\ref{eq:EClingamma}


As shown in Refs.~\cite{Stalhammar-NJP2023,
MunozArboleda2026ThermodynamicsAnalogue} and reproduced for the open fermionic system in Sec.~\ref{sec:OpenFermionicsystems}, the tilted exceptional cone defines an effective $1+1$-dimensional metric Eq.~\ref{eq:PGmetric}, which corresponds to the Painlevé-Gullstrand Schwarzschild black-hole spacetime. The profile in Eq.~\eqref{eq:horizonprofile} produces interfaces at
which $|f(r)|=1$; these define analogue horizons.  Reversing
$f(r)$ changes the sign of the off-diagonal metric component and
interchanges the black-hole and white-hole orientations.  This is the
single-particle interpretation of the two rapidity sectors found in
Sec.~\ref{sec:OpenFermionicsystems}.

\begin{figure}
\centering
\includegraphics[width=\textwidth]{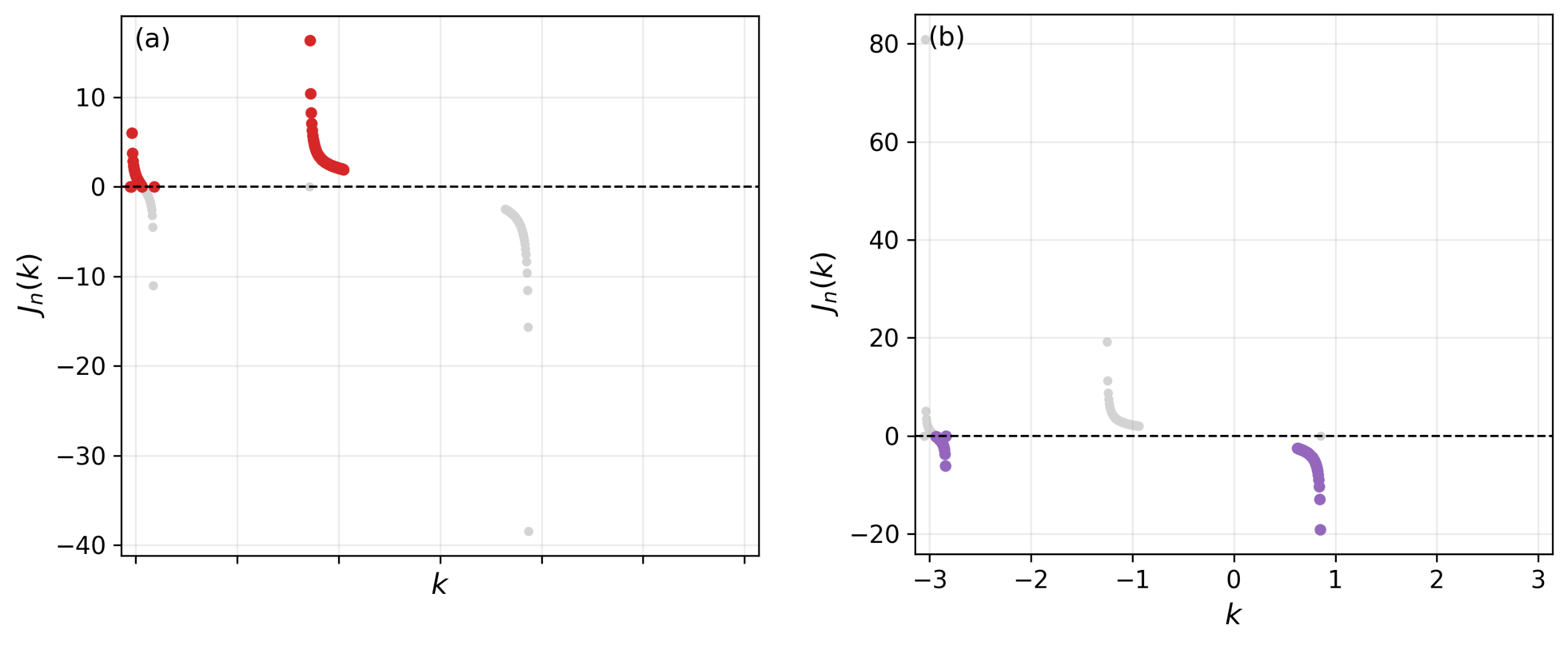}
\caption{Biorthogonal flux of the asymptotic Bloch modes. Gray points denote weakly damped positive $\mathrm{Re}[\,E_n(k)]$ modes. 
(a) Exterior ($\kappa=\kappa_2$); (b) interior ($\kappa=\kappa_1$) fluxes in the asymptotic region. Modes with $J_n(k)>0$ [$J_n(k)<0$] are highlighted in red (purple). Calculations were made with $\tau=1$, $\gamma=0.3$, $\kappa_1=1.01$, and $\kappa_2=0.99$.
}
\label{fig:Flux}
\end{figure}

For a uniform asymptotic chain (far away from the horizon) and for each $k$, let $\ket{\psi_{R,n}(k)}$ and $\ket{\psi_{L,n}(k)}$ be the right and left eigenvectors of $\mathcal{H}_{\text{B}}(k)$,
\begin{align}
	\mathcal{H}_{\text{B}}(k)\ket{\psi_{R,n}(k)}&=E_n(k)\ket{\psi_{R,n}(k)},\nonumber \\
	\bra{\psi_{L,n}(k)} \mathcal{H}_{\text{B}}(k)&=E_n(k)\bra{\psi_{L,n}(k)},
\end{align}
which are biorthogonally normalized as
\begin{equation}
	\braket{\psi_{L,m}(k)|\psi_{R,n}(k)}=\delta_{mn}.
\end{equation}
The group-velocity (flux) is
\begin{equation}
	J_n(k)=\Re\!\left[ \bra{\psi_{L,n}(k)}v(k)\ket{\psi_{R,n}(k)} \right],
\end{equation}
with $v(k)=\partial \mathcal{H}_{\rm B}(k)/\partial k$. The role of $J_n(k)$ is to assign a propagation direction to each asymptotic Bloch mode in a non-Hermitian setting. For the horizon at $r=r_1$, the exterior and interior regions lie
on opposite sides of the interface.  With the orientation adopted in
Fig.~\ref{fig:Flux}, modes with $J_n(k)>0$ in the exterior and
$J_n(k)<0$ in the interior propagate away from the horizon and
define the outgoing Hawking-like and partner-like channels. This is essential in the next section, where the Hawking signature is built from the anomalous correlation between the outgoing Hawking-like exterior channel and the outgoing interior partner channel.

\section{Many-body effective model and analogue Hawking-radiation witnesses}
\label{sec:many_body_witnesses}

Sections~\ref{sec:OpenFermionicsystems} and \ref{sec:SPNHTBH} established the open-system and kinematic foundations of the analogue-horizon problem. 
These results, however, concern spectral and one-body observables and do not by themselves establish Hawking pair production. We therefore introduce an effective fermionic many-body extension and formulate the stationary fixed-frequency scattering problem in terms of particle-like and hole-like asymptotic channels, which allows us to define the outgoing Hawking and partner moments (Sec.~\ref{subsec:fermionic_nambu_channels}). Then, we reconstruct the channel correlations in real space and use the anomalous density-density contribution to identify the cross-horizon Hawking signatures and compare them with the control profiles (Sec.~\ref{subsec:nonlocal_correlations_F}). Finally, we analyze the frequency-resolved channel populations and anomalous pair amplitude and test their consistency with covariance positivity and Pauli exclusion in Sec.~\ref{subsec:fermionic_physicality}.

\subsection{Fermionic Nambu extension and scattering channels}
\label{subsec:fermionic_nambu_channels}

The non-Hermitian GL-NNN Hamiltonian introduced in Sec.~\ref{sec:OpenFermionicsystems} is number conserving. Its single-particle equations therefore couple annihilation operators among themselves and do not mix them with creation operators. Consequently, the GL-NNN Hamiltonian preserves the global $U(1)$ structure and cannot by itself generate the anomalous covariance required to describe particle-partner pair production.

To fix the notation from the outset, we collect the two physical
fermionic annihilation operators in unit cell $j$ into the sublattice
spinor
\begin{equation}
 \hat{\bm\Psi}_j= \left(\hat a_j,\hat b_j\right)^T, \hspace{0.2cm}
 \hat{\bm\Psi}= \left(\hat{\bm\Psi}_0, \hat{\bm\Psi}_1, ...\hat{\bm\Psi}_{N-1}\right)^T.
 \label{eq:secIV_Psi_def_F}
\end{equation}
Here, $\hat{\bm\Psi}_j$ acts in the physical $A$-$B$ sublattice
space and should not be confused with the Majorana operators
$c_{j,s}$ and $d_{j,s}$ introduced in
Sec.~\ref{sec:OpenFermionicsystems}. 
The corresponding local Nambu operator in cell $j$ is
\begin{equation}
 \hat{\bm\Upsilon}_j= \left(\hat{\bm\Psi}_j,\hat{\bm\Psi}_j^{\dagger T}\right)^T
 = \left(\hat a_j,\hat b_j,\hat a_j^\dagger,\hat b_j^\dagger\right)^T.
 \label{eq:secIV_local_Nambu_F}
\end{equation}

Introducing the Nambu basis allows us to extend the original number-conserving model by coupling its particle and hole sectors. We therefore consider the minimal quadratic anomalous extension of the GL-NNN Hamiltonian. 
This extension breaks the global $U(1)$ particle-number symmetry while preserving fermion parity and provides the particle-hole mixing required to construct Hawking-partner correlations. For spinless fermions, the pairing matrix must be antisymmetric $\Delta_{\rm F}
 =\delta_{\rm F}(i\sigma^y)\otimes\mathbb{I}_N$, $\Delta_{\rm F}^T=-\Delta_{\rm F}$, where $\delta_{\rm F}$ is a real intracell $A$-$B$ pairing amplitude, and we choose the simplest spatially uniform pairing as,
\begin{equation}
 \hat H_{\rm pair}^{\rm F}
 =\frac12\left(
 \hat{\bm\Psi}^{\dagger}\Delta_{\rm F}
 \hat{\bm\Psi}^{\dagger T}
 +{\rm h.c.}\right).
 \label{eq:secIV_pair_F}
\end{equation}
The pairing term is introduced here as an effective Gaussian
many-body extension and is not derived from the microscopic
Lindblad construction of Sec.~\ref{sec:OpenFermionicsystems}. It is spatially uniform and is kept identical in the simulations performed in this section. Hence, neither the position nor the slope of the final correlation ridge is inserted through $\Delta_{\rm F}$.

Because the background profile $f(r)$ is time independent, the stationary scattering problem is invariant under time translations and the frequency $\omega$ of the Nambu excitations is conserved across the interface. We therefore resolve the Nambu field into components proportional to $e^{-i\omega t}$ and formulate the scattering problem independently at each real $\omega$. In a uniform asymptotic region $\mathscr{r}$, the allowed spatial modes are obtained from
\begin{equation}
 \det\!\left[
 \mathcal K_{\rm F}(q;f_\mathscr{r})
 -\omega\mathbb{I}_4
 \right]=0,
 \label{eq:secIV_dispersion_fixed_omega_F}
\end{equation}
where $\mathcal K_{\rm F}(q;f_\mathscr{r})$ is the fermionic Nambu-Bloch matrix defined explicitly in Appendix~\ref{app:recurrence_matrices_v8}. Equation~\eqref{eq:secIV_dispersion_fixed_omega_F} provides a fixed-frequency formulation of the usual band-structure problem: rather than fixing $q$ and calculating a dispersion $E_n(q)$, we fix $\omega$ and determine all wave numbers $q_{\mathscr{r}\mu\sigma}(\omega)$ satisfying $E_n(q_{\mathscr{r}\mu\sigma})=\omega$. From now on, $\mu$ distinguishes the different wave-number solutions, and $\sigma\in\{+,-\}$ denotes the particle-like or hole-like Nambu sector. Since we set $\hbar=1$, energy and frequency have the same units, and the numerical results below are expressed in terms of $\omega/\tau$.

Non-Hermiticity appears in the spatial solutions, for which the roots $q_{\mathscr{r}\mu\sigma}(\omega)$ may be complex. Writing $q=q_{\rm R}+iq_{\rm I}$ gives $e^{iqj}=e^{iq_{\rm R}j}e^{-q_{\rm I}j}$, so that ${\rm Im}\,q$ describes spatial growth or decay. Such decaying modes are retained in the interface matching but do not define independent asymptotic incoming or outgoing channels. For each propagating root, the corresponding Nambu eigenvector and biorthogonal flux determine the particle-like or hole-like character and the propagation direction (see Appendix \ref{app:asymptotic_modes_consistent}). These propagating solutions provide the asymptotic basis for a fixed-frequency scattering description. Once their corresponding
channel amplitudes are introduced below and classified as incoming or
outgoing, the scattering through the inhomogeneous horizon region can
be encoded in a scattering matrix $\mathcal S(\omega)$, which relates
the incoming and outgoing channels at the same conserved frequency.

In each uniform asymptotic region, the stationary Nambu problem at fixed $\omega$ therefore has particle-like and hole-like solutions. We denote the corresponding right Nambu eigenmodes by $\bm\psi^R_{\mathscr{r}\mu\sigma}(j,\omega)$, in terms of which an expansion of the field is written as
\begin{align}
 \hat{\bm\Upsilon}_{\mathscr{r}}(j,\omega)
 =
 &\sum_{\mu\in +}
 \bm\psi^R_{\mathscr{r}\mu+}(j,\omega)\,
 \hat\eta_{\mathscr{r}\mu+}(\omega) + \sum_{\mu\in -}
 \bm\psi^R_{\mathscr{r}\mu-}(j,\omega)\,
 \hat\eta_{\mathscr{r}\mu-}^{\dagger}(\omega).
 \label{eq:secIV_eta_mode_expansion_F}
\end{align}
Thus, the Nambu eigenmodes $\bm\psi^R$ are c-number mode functions, whereas the $\hat\eta$'s are their annihilation or creation operator amplitudes. Equivalently, the channel operators are obtained by projecting the lattice Nambu field onto the corresponding dual left eigenmodes, as shown explicitly in Appendix~\ref{app:channel_operators_v9}. 

We evaluate the stationary Gaussian problem in the frequency domain by matching the finite inhomogeneous region to these asymptotic channels. Collecting all propagating channels gives the multichannel scattering relation
\begin{equation}
 \begin{pmatrix}
  \hat{\bm\eta}_{+,{\rm out}}\\
  \hat{\bm\eta}_{-,{\rm out}}^\dagger
 \end{pmatrix}
 =
 \begin{pmatrix}
  \mathcal S_{++}&\mathcal S_{+-}\\
  \mathcal S_{-+}&\mathcal S_{--}
 \end{pmatrix}
 \begin{pmatrix}
  \hat{\bm\eta}_{+,{\rm in}}\\
  \hat{\bm\eta}_{-,{\rm in}}^\dagger
 \end{pmatrix}.
 \label{eq:secIV_S_F}
\end{equation}
The two indices of each scattering block $\mathcal S_{\sigma_{\rm out}\sigma_{\rm in}}$ label the Nambu sector of the outgoing and incoming channels, respectively. Thus, $\mathcal S_{+-}$ maps an incoming hole-like channel to an outgoing particle-like channel, whereas $\mathcal S_{-+}$ maps an incoming particle-like channel to an outgoing hole-like channel. The diagonal blocks describe scattering within the same Nambu sector. The incoming state is chosen as the channel vacuum. In the special canonical two-channel limit, Eq.~\eqref{eq:secIV_S_F} reduces to a fermionic Bogoliubov transformation with coefficients $\mathcal U_\omega$ and $\mathcal V_\omega$ satisfying $|\mathcal U_\omega|^2+|\mathcal V_\omega|^2=1$. The conditional non-Hermitian scattering matrix used here does not need to satisfy this canonical relation exactly. The explicit recurrence, channel construction, and mode-matching equations are given in Appendices~\ref{app:recurrence_matrices_v8} and \ref{app:interface_scattering_v8}.

From the full outgoing channel vectors, we select the exterior Hawking channel and the interior partner channel,
\begin{equation}
 \hat\eta_H(\omega)
 \equiv
 \hat\eta_{\mathscr{r}_H\mu_H+,{\rm out}}(\omega),
 \hspace{0.2cm}
 \hat\eta_P(\omega)
 \equiv
 \hat\eta_{\mathscr{r}_P\mu_P-,{\rm out}}(\omega).
 \label{eq:secIV_selected_eta_F}
\end{equation}
Their frequency-resolved moments are
\begin{align}
 n_H(\omega)&=\langle\hat\eta_H^\dagger\hat\eta_H\rangle, \hspace{0.2cm}
 n_P(\omega)=\langle\hat\eta_P^\dagger\hat\eta_P\rangle, \hspace{0.2cm}
 m_{HP}(\omega)=\langle\hat\eta_H\hat\eta_P\rangle.
 \label{eq:secIV_channel_moments_F}
\end{align}
For the incoming channel vacuum (see Appendices \ref{app:interface_scattering_v8}, \ref{app:finite_matching_consistent} and \ref{app:incoming_vacuum_consistent}), Eq.~\eqref{eq:secIV_S_F} gives
\begin{align}
 n_H(\omega)
 &=\left[\mathcal S_{+-}\mathcal S_{+-}^\dagger\right]_{HH},
 \\
 n_P(\omega)
 &=\left[\mathcal S_{-+}\mathcal S_{-+}^\dagger\right]_{PP},
 \\
 m_{HP}(\omega)
 &=\left[\mathcal S_{++}\mathcal S_{-+}^\dagger\right]_{HP}.
 \label{eq:secIV_moments_blocks_F}
\end{align}
Here, $n_H$ and $n_P$ are occupations of asymptotic scattering channels. The two descriptions are related through the asymptotic mode expansion in Eq.~\eqref{eq:secIV_eta_mode_expansion_F}.

\subsection{Fermionic Hawking witnesses and non-local density correlations}
\label{subsec:nonlocal_correlations_F}

The most direct real-space signature of an analogue Hawking radiation is the connected density-density correlator
\begin{equation}
 G^{(2)}_{pq,{\rm conn}}
 =
 \left\langle
 \delta\hat n_p\,\delta\hat n_q
 \right\rangle,
 \hspace{0.2cm}
 \delta\hat n_p=\hat n_p-\langle\hat n_p\rangle,
 \label{eq:secIV_G2_def_F}
\end{equation}
where $p=(m,s)$ and $q=(n,s')$ denote distinct lattice modes and $\hat n_p=\hat\Psi_p^\dagger\hat\Psi_p$ is the local number operator. A long-range ridge joining cells on opposite sides of the horizon is the lattice analogue of the Hawking ``moustache'' predicted for acoustic black holes \cite{BalbinotEtAl2008,CarusottoEtAl2008NJP}.

The channel operator $\hat\eta$ and the real-space lattice operators contained in $\hat{\bm\Psi}$ describe the same fermionic field in different bases. For the selected outgoing channels, the annihilation sector of Eq.~\eqref{eq:secIV_eta_mode_expansion_F} may be written in the asymptotic windows as
\begin{align}
 \hat\Psi_{(n,s)}(\omega)
 &\supset
 \varphi_{H,s}(n,\omega)\hat\eta_H(\omega),
 \qquad n>r_1,
 \nonumber\\
 \hat\Psi_{(m,s)}(\omega)
 &\supset
 \varphi_{P,s}(m,\omega)\hat\eta_P(\omega),
 \qquad m<r_1.
 \label{eq:secIV_varphi_eta_relation_F}
\end{align}
The quantities $\varphi_{H,s}$ and $\varphi_{P,s}$ are therefore Hawking and partner c-number annihilation-sector spatial envelopes extracted from the corresponding Nambu eigenmodes $\bm\psi^R$. They contain the asymptotic wave numbers, flux normalization, and sublattice spinors of the selected channels. Their explicit forms are given in Appendix~\ref{app:covariance_wick_v8}.

Using Eq.~\eqref{eq:secIV_varphi_eta_relation_F}, the anomalous Hawking-partner channel moment is converted back to the lattice basis as
\begin{equation}
 F^{HP}_{ms,ns'}
 =
 \int_{\omega_a}^{\omega_b}\frac{d\omega}{2\pi}\,
 m_{HP}(\omega)\,
 \varphi_{P,s}(m,\omega)
 \varphi_{H,s'}(n,\omega).
 \label{eq:secIV_F_reconstruction_F}
\end{equation}
This equation makes the chain of definitions explicit: $m_{HP}$ is a correlation between the channel operators $\hat\eta_H$ and $\hat\eta_P$, while the envelopes $\varphi_H$ and $\varphi_P$ map that channel correlation back to the original lattice operators $\hat a_j$ and $\hat b_j$.

We use the components $\hat\Psi_p$ of the physical lattice spinor consistently to define the normal (number conserving) and anomalous (pairing) covariances,
\begin{equation}
 \Gamma_{pq}
 =
 \langle\hat\Psi_q^\dagger\hat\Psi_p\rangle,
 \hspace{0.2cm}
 F_{pq}
 =
 \langle\hat\Psi_p\hat\Psi_q\rangle,
 \hspace{0.2cm}
 \bar F_{pq}
 =
 \langle\hat\Psi_p^\dagger\hat\Psi_q^\dagger\rangle.
 \label{eq:secIV_covariances_F}
\end{equation}
The normal covariance $\Gamma_{pq}$ describes single-particle
occupations and coherences, whereas $F_{pq}$ and $\bar F_{pq}$
measure pair coherence between two annihilation or two creation
operators. The anomalous covariances are enabled by the
particle-hole mixing introduced through the Nambu extension and are
therefore the quantities that encode the real-space pairing
correlations relevant for Hawking-partner production. For fermions, $F^{T}=-F$ and $\bar F=F^\dagger$. For two distinct lattice modes $p\neq q$, Wick's theorem gives
\begin{equation}
 G_{pq,{\rm conn}}^{(2),{\rm F}}
 =
 -|\Gamma_{pq}|^2+|F_{pq}|^2.
 \label{eq:secIV_WickF_only}
\end{equation}
The complete contraction, including the fermionic sign, is derived in Appendix~\ref{app:covariance_wick_v8}.

Figure~\ref{fig:secIV_cross_F} displays the positive anomalous contribution
\begin{equation}
 \mathcal P_{mn}^{\rm F}
 =
 \sum_{s,s'\in\{A,B\}}
 \left|F^{\rm F}_{ms,ns'}\right|^2.
 \label{eq:secIV_pair_map_F}
\end{equation}


To determine whether the reconstructed correlations are specifically
organized by the analogue horizon, rather than being generated by the
pairing term or by the spatial inhomogeneity alone, we perform three
calculations using the same pairing amplitude, frequency grid, and reconstruction procedure. The horizon profile is created at the critical value $f=\kappa_3=1$ for $\kappa_1=1.2$ and $\kappa_2=0.8$. The no-crossing control retains the same two smooth interfaces at $r_1$ and $r_2$, but uses $\kappa_1^{\rm nc}=0.98$ and $\kappa_2^{\rm nc}=0.94$, so that the profile remains below the critical value everywhere. Finally, the uniform control is established for $f(j)=1.2$ throughout the chain and therefore contains neither an interface nor a horizon. This comparison allows us to distinguish horizon-induced spatial organization from correlations produced by the pairing term or by the presence of an interface alone.

For the horizon profile, at the first interface the outgoing Hawking
mode is reconstructed to the right of $r_1$, while its partner is
reconstructed to the left. Their distances from the interface are
$d_H=j-r_1>0$ and $d_P=r_1-i>0$. If the two excitations are emitted
simultaneously, the dominant correlation is expected to satisfy
$d_H/|v_H(\omega)|\simeq d_P/|v_P(\omega)|$, where $v_H$ and $v_P$
are the biorthogonal flux velocities of the selected outgoing channels.
The cyan dashed line in the cross-interface panels follows this
independently calculated velocity ratio and is therefore not fitted to
the correlation map.

\begin{figure}
 \centering
 \includegraphics[width=\columnwidth]{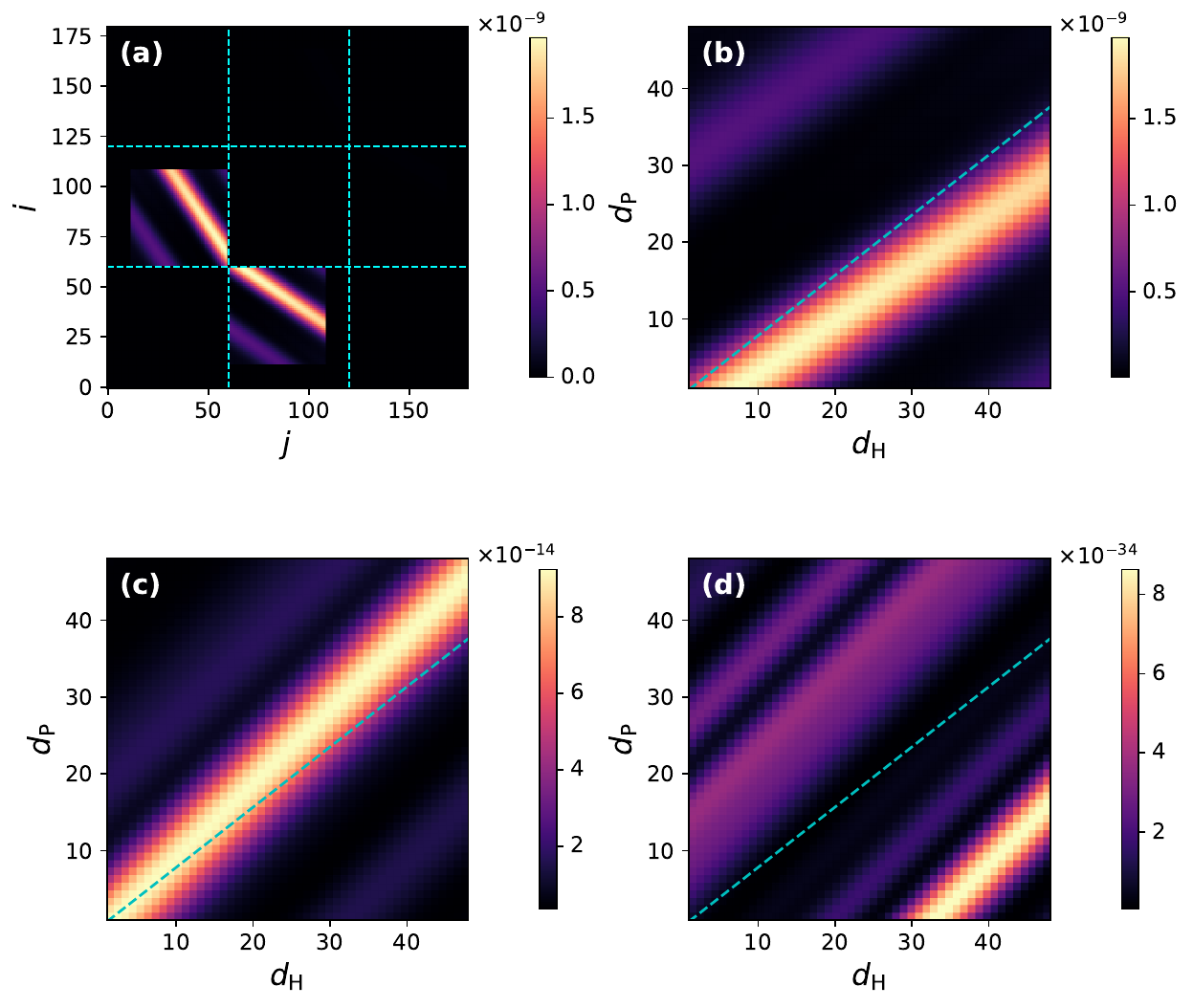}
 \caption{Frequency-integrated anomalous Hawking-partner correlations for the fermionic Nambu extension. (a) The horizon calculation in the original $N\times N$ unit-cell coordinate plane $(i,j)$. The scattering calculation reconstructs finite windows of $D=48$ cells on both sides of each interface and embeds the calculated blocks at their physical lattice positions. The cyan horizontal and vertical dashed lines mark $r_1$ and $r_2$. (b) The isolated cross-interface block around $r_1$ for the horizon profile with $\kappa_1=1.2$ and $\kappa_2=0.8$. (c) The no-crossing control with $\kappa_1=0.98$ and $\kappa_2=0.94$, and (d) the uniform control with $\kappa_1=\kappa_2=1.2$. The cyan diagonal in (b) to (d) follows the independently calculated equal-travel-time relation. For all figures $N=180$, $r_1=60$, $r_2=120$, $\tau=1$, $\gamma=0.5$, $l=1.5$, and $\delta_{\rm F}=0.2$; 61 frequencies are integrated over $0.06\leq\omega/\tau\leq0.95$.}
 \label{fig:secIV_cross_F}
\end{figure}

The map in Fig.~\ref{fig:secIV_cross_F}(a) shows two clear branches anchored at the two interfaces required by PBC. Figure~\ref{fig:secIV_cross_F}(b) isolates the branch associated with the first black-hole-oriented interface and shows that its direction is consistent with the outgoing-channel velocities determined in Sec.~\ref{sec:SPNHTBH}. Figures~\ref{fig:secIV_cross_F}(c) and (d) show non-crossing control and uniform control setups, respectively,  where no interface or horizon is obtained. The same observable is suppressed by more than four orders of magnitude in the no-crossing interface and is reduced to numerical round-off in the uniform chain. These controls establish a horizon-organized anomalous Hawking-partner correlation in the fermionic extension which provides a fermionic analogue of Hawking radiation.

\subsection{Fermionic covariance-positivity criterion}
\label{subsec:fermionic_physicality}

The separate quantity $|m_{HP}(\omega)|$ measures the coherent pair amplitude entering Eq.~\eqref{eq:secIV_F_reconstruction_F}. The analysis window $0.06\leq\omega/\tau\leq0.95$ follows the first smooth low-energy Hawking-partner branch, avoids the flux-normalization problem very close to zero frequency, and excludes higher-frequency channel rearrangements. The broad scan and the numerical selection are documented in Appendix~\ref{app:frequency_window_v8}. To quantify the suppression of the selected channel in the controls, we define
\begin{equation}
 \mathcal I_H^X
 =
 \int_{\omega_a}^{\omega_b}
 d\omega\,n_H^X(\omega),
 \qquad
 \mathcal E_H
 =
 \frac{\mathcal I_H^{\rm horizon}}
 {\mathcal I_H^{\rm no\text{-}crossing}},
 \label{eq:secIV_enhancement_F}
\end{equation}
where $X\in\{{\rm horizon},{\rm no\text{-}crossing},{\rm uniform}\}$. The enhancement factor $\mathcal E_H$ measures horizon selectivity of the chosen output branch. Positivity of the fermionic Gaussian covariance implies
\begin{equation}
 |m_{HP}|^2
 \leq
 \min\!\left[
 n_H(1-n_P),
 (1-n_H)n_P
 \right].
 \label{eq:secIV_Pauli_F}
\end{equation}
The corresponding normalized Pauli ratio is
\begin{equation}
 \Theta_{HP}^{\rm F}
 =
 \frac{|m_{HP}|}
 {\sqrt{\min\!\left[
 n_H(1-n_P),
 (1-n_H)n_P
 \right]}}
 \leq1.
 \label{eq:secIV_thetaF_only}
\end{equation}
The condition $\Theta_{HP}^{\rm F}\leq1$ is required by fermionic covariance positivity and Pauli exclusion. Values below one show that the anomalous covariance is compatible with a physical fermionic state, while $\Theta_{HP}^{\rm F}>1$ would indicate an unphysical covariance or a failure of the conditional map to preserve the canonical anticommutation relations. The ratio is therefore used here as a consistency criterion rather than as evidence of fermionic nonclassicality \cite{Bravyi2005}.

\begin{figure}
 \centering
 \includegraphics[width=0.97\textwidth]{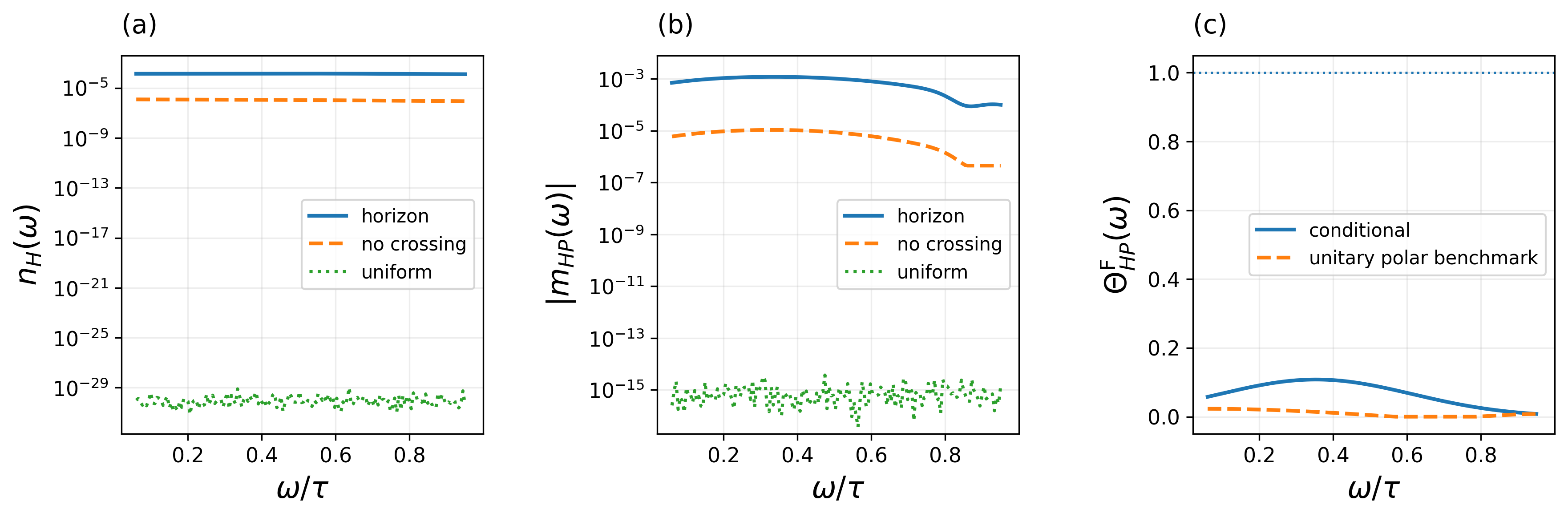}
 \caption{Frequency-resolved fermionic channel diagnostics. (a) The conversion $n_H(\omega)$ into the selected outgoing Hawking channel for the horizon, no-crossing, and uniform profiles. (b) The coherent anomalous amplitude $|m_{HP}(\omega)|$ for the same profiles. (c) The normalized fermionic Pauli ratio. The solid curve is calculated from the conditional non-Hermitian scattering matrix, while the dashed curve is the unitary polar benchmark defined in Appendix~\ref{app:canonical_completion_v8}. The horizontal dotted line denotes 1. Over $0.06\leq\omega/\tau\leq0.95$, the integrated conversion weights are $\mathcal I_H^{\rm hor}=1.223\times10^{-4}$ and $\mathcal I_H^{\rm nc}=9.456\times10^{-7}$, giving $\mathcal E_H=129.4$; the uniform value is of order $10^{-30}$. The maximal conditional ratio is $0.108$, while the unitary polar benchmark reaches $0.023$. Parameters are the same as in Fig.~\ref{fig:secIV_cross_F}.}
 \label{fig:secIV_frequency_F}
\end{figure}

Figure~\ref{fig:secIV_frequency_F} shows the frequency-resolved
channel quantities underlying the real-space Hawking-partner
correlations. In Fig.~\ref{fig:secIV_frequency_F}(a) the outgoing Hawking channel for the horizon, non-crossing and uniform profiles are depicted and in Fig.~\ref{fig:secIV_frequency_F}(b) the coherent anomalous amplitude are shown for the same profiles. The selected Hawking output is populated strongly only when the profile crosses the horizon condition, and the same frequency region carries the coherent Hawking-partner moment used to construct Fig.~\ref{fig:secIV_cross_F}. In Fig.~\ref{fig:secIV_frequency_F}(c) we show that the fermionic correlation remains below the Pauli bound both conditionally and after the unitary polar benchmark. These results establish horizon-selective fermionic Hawking-partner conversion with a physical anomalous covariance. 


\section{Discussion \& Conclusions}
\label{sec:concandout}

\subsection{Summary}

In this work, we studied a microscopic Markovian open fermionic system and investigated how analogue black-hole physics emerges within its effective non-Hermitian description. The conditional no-jump evolution generates, up to an overall imaginary shift, GL and non-reciprocal NNN terms of the tight-binding model coupled to a reservoir. Through third quantization, the damping matrix separates into two sectors related by reversal of the non-reciprocal hopping, which can be associated with analogue black-hole and white-hole configurations. Their rapidity spectra exhibit tilted exceptional cones, while the equal-momentum particle densities and steady-state current display sharp variations at the two interfaces, showing that the horizon structure survives the long-time approach to the nonequilibrium steady state.

At the effective single-particle level, the non-reciprocal NNN hopping controls the tilt of the exceptional cones and plays the role of an effective flow field. Its kink profile produces a Painlevé-Gullstrand geometry, with interfaces at which the effective flow crosses the analogue propagation speed. Because the Hamiltonian is non-Hermitian, propagation directions are identified through the biorthogonal flux rather than through an ordinary Hermitian group velocity. This construction selects outgoing exterior and interior channels associated with the first horizon, providing the single-particle basis for the subsequent correlation analysis.

The existence of a horizon and the observation of one-body steady-state signatures are not sufficient to establish Hawking pair production. We therefore introduced a fermionic Gaussian Nambu extension with a weak spatially uniform anomalous coupling, chosen antisymmetric as required for spinless fermions. The resulting scattering matrix determines the Hawking and partner populations and their anomalous correlation, which was used to reconstruct the non-local covariance in real space. The correlations form long-range branches anchored at the two interfaces and, around the first horizon, follow the equal-travel-time trajectory determined independently from the outgoing channel velocities. Their strong suppression in the no-crossing profile and reduction to numerical roundoff in the uniform system show that they are not generated by the uniform pairing term alone. Together with the enhanced frequency-resolved conversion and anomalous amplitude, these results establish horizon-organized fermionic Hawking-partner production and the characteristic Hawking-moustache structure within the effective non-Hermitian theory
\cite{BalbinotEtAl2008,CarusottoEtAl2008NJP,MacherParentani2009,RecatiPavloffCarusotto2009}.

\subsection{Discussion and outlook}

The fermionic frequency-resolved ratio has a different role from a bosonic nonclassicality criterion. The condition $\Theta_{HP}^{\rm F}\leq1$ is imposed by covariance positivity and Pauli exclusion and therefore constitutes a physicality requirement rather than a nonclassicality witness. Our calculation consequently establishes horizon-selective particle-partner conversion together with a physical anomalous covariance, while its genuinely quantum character must be assessed through a fermionic entanglement criterion
\cite{ShapourianShiozakiRyu2017,ShapourianRyu2019}. The fermionic Gaussian Nambu extension should be regarded as a controlled phenomenological construction that reveals how the single-particle analogue geometry organizes pair correlations once an anomalous sector is present.

A natural next step is to derive the anomalous sector directly from a microscopic open-system realization. Such a construction would determine whether the particle-hole mixing introduced phenomenologically in the present Nambu extension can arise from additional coherent or dissipative processes compatible with the Lindblad dynamics. A positive fermionic logarithmic negativity, calculated using a fermionic partial transpose or partial-time-reversal construction, would instead demonstrate that the Hawking and partner channels form a genuinely entangled pair
\cite{ShapourianShiozakiRyu2017,ShapourianRyu2019}. Combining such an entanglement calculation with the no-crossing and uniform controls would distinguish horizon-induced quantum pairing from correlations generated by the anomalous term or by the reservoirs alone.

The microscopic open-system construction may be approached experimentally
using ultracold atoms in optical lattices
\cite{MiyakeEtAl2013,GouEtAl2020,LiangEtAl2022,RenEtAl2022,ZhaoEtAl2025}
or synthetic momentum-space lattices \cite{LappEtAl2019}, while the
effective non-Hermitian Hamiltonian may also be realized in photonic arrays
\cite{WeidemannEtAl2020,SlootmanCherifi2024}, synthetic frequency lattices
\cite{ChengEtAl2023}, topolectrical circuits
\cite{HelbigHofman2020}, or programmable superconducting processors
\cite{ShiEtAl2023}. The Hawking moustaches can be measured by repeatedly
preparing the system and reconstructing the connected density covariance
between lattice cells on opposite sides of the interface
\cite{steinhauer_observation_2016,deNova2019}. For the fermionic system
considered here, reconstruction of the reduced two-channel covariance would
provide direct access to the Hawking and partner occupations and anomalous
moment and would allow a fermionic entanglement negativity to be evaluated
\cite{ShapourianShiozakiRyu2017,ShapourianRyu2019}.

The framework also permits a systematic study of finite-size effects, disorder, boundary conditions, imperfect GL balance, and the width of the spatial horizon profile. It provides a direct setting in which to determine whether the anomalous channel spectrum contains a thermal interval governed by the analogue surface gravity and how this scale is related to the thermodynamic signatures obtained previously for the effective single-particle model
\cite{MunozArboleda2026ThermodynamicsAnalogue}.

\section*{Acknowledgements}
DFM-A acknowledges funding from the Colombian Ministry of Science (Minciencias) and CMS acknowledges funding from QuMat, a program of the Netherlands organization for Scientific Research (NWO) that is funded by the Dutch Ministry of Education, Culture and Science. M.S. is supported by the Swedish Research Council (VR) under Grant No. 2024.00272. 



\begin{appendix}
\numberwithin{equation}{section}

\section{Stationary real-space recurrence}
\label{app:recurrence_matrices_v8}

This appendix provides the technical construction underlying the
fermionic scattering analysis of Sec.~\ref{sec:many_body_witnesses}.
We use the notation introduced in the main text:
$\hat{\bm\Psi}_j=(\hat a_j,\hat b_j)^{\mathsf T}$ denotes the
two-component physical annihilation spinor in cell $j$, while
$\hat{\bm\Upsilon}_j=(\hat a_j,\hat b_j,\hat a_j^\dagger,\hat b_j^\dagger)^{\mathsf T}$
denotes the corresponding local Nambu operator. The superscripts
$R$ and $L$ always denote right and left eigenvectors of a
non-Hermitian Nambu-Bloch matrix; propagation direction is determined
instead by the biorthogonal flux.


We first derive the c-number stationary equations that determine the
Nambu modes.  To distinguish a mode amplitude from the operator
Nambu field $\hat{\bm\Upsilon}_j$, we denote the stationary
four-component amplitude in cell $j$ by
\begin{equation}
 \bm\zeta_j(\omega)
 =
 \left(
 u_{j,A},
 u_{j,B},
 v_{j,A},
 v_{j,B}
 \right)^{T}.
 \label{eq:app_stationary_mode_amplitude_consistent}
\end{equation}
The onsite normal block is
\begin{equation}
 h_0=
 \begin{pmatrix}
 i\gamma&\tau\\
 \tau&-i\gamma
 \end{pmatrix},
 \label{eq:app_h0_consistent}
\end{equation}
and, consistently with Sec.~\ref{subsec:fermionic_nambu_channels},
the local $A$-$B$ block of the full pairing matrix is
\begin{equation}
 \Delta_{{\rm F},0}=\delta_{\rm F}i\sigma^y,
 \qquad
 \Delta_{{\rm F},0}^{T}=-\Delta_{{\rm F},0},
 \label{eq:app_Delta_F_consistent}
\end{equation}
so that the full matrix used in the main text is
$\Delta_{\rm F}=\Delta_{{\rm F},0}\otimes\mathbb{I}_N$.
The corresponding onsite Nambu matrix is
\begin{equation}
 V_{\rm F}=
 \begin{pmatrix}
 h_0&\Delta_{{\rm F},0}\\
 -\Delta_{{\rm F},0}^*&-h_0^*
 \end{pmatrix}.
 \label{eq:app_V_v8}
\end{equation}

For a bond with non-reciprocal coefficient $f$, the normal forward
and backward hopping blocks are
\begin{equation}
 t_+(f)=
 \begin{pmatrix}
 f/2&0\\
 \tau&-f/2
 \end{pmatrix},
 \qquad
 t_-(f)=
 \begin{pmatrix}
 -f/2&\tau\\
 0&f/2
 \end{pmatrix},
 \label{eq:app_tpm_consistent}
\end{equation}
with Nambu extensions
\begin{equation}
 T_\pm(f)=
 \begin{pmatrix}
 t_\pm(f)&0\\
 0&-t_\pm^*(f)
 \end{pmatrix}.
 \label{eq:app_Tpm_v8}
\end{equation}
If $f_{j+1/2}$ is the coefficient on the bond connecting cells $j$
and $j+1$, the stationary equation at real frequency $\omega$ is
\begin{equation}
 T_+(f_{j+1/2})\bm\zeta_{j+1}
 +\left[V_{\rm F}-\omega\mathbb{I}_4\right]\bm\zeta_j
 +T_-(f_{j-1/2})\bm\zeta_{j-1}=0.
 \label{eq:app_recurrence_v8}
\end{equation}
Thus, $V_{\rm F}$ contains the intracell hopping, GL term, and pairing,
while $T_+$ and $T_-$ contain the forward and backward intercell
hopping.

In a uniform asymptotic region with constant $f_\mathscr{r}$, the
corresponding Nambu-Bloch matrix is
\begin{equation}
 \mathcal K_{\rm F}(q;f_\mathscr{r})
 =
 V_{\rm F}
 +T_+(f_\mathscr{r})e^{iq}
 +T_-(f_\mathscr{r})e^{-iq}.
 \label{eq:app_Kq_v8}
\end{equation}
At fixed real $\omega$, the allowed wave numbers are the roots of
\begin{equation}
 \det\!\left[
 \mathcal K_{\rm F}(q;f_\mathscr{r})
 -\omega\mathbb{I}_4
 \right]=0.
 \label{eq:app_fixed_frequency_roots}
\end{equation}
This is the fixed-frequency problem described in
Sec.~\ref{subsec:fermionic_nambu_channels}: $\omega$ is held fixed
and the spatial roots $q_{\mathscr{r}\mu\sigma}(\omega)$ are determined.
The index $\mu$ distinguishes the different wave-number solutions
within a given asymptotic region and Nambu sector.

The finite inhomogeneous region is not propagated by multiplying long
transfer matrices.  Instead, the block-tridiagonal equations of all
interface cells are solved simultaneously with the asymptotic
boundary conditions.  This avoids numerical instabilities associated
with products of exponentially growing and decaying transfer
eigenvalues.

\section{Asymptotic Nambu modes and biorthogonal channel classification}
\label{app:asymptotic_modes_consistent}

For each root $q_{\mathscr{r}\mu\sigma}(\omega)$, the right and left
eigenvectors satisfy
\begin{align}
 \mathcal K_{\rm F}(q_{\mathscr{r}\mu\sigma};f_\mathscr{r})
 \ket{\psi^R_{\mathscr{r}\mu\sigma}}
 &=
 \omega\ket{\psi^R_{\mathscr{r}\mu\sigma}},
 \nonumber\\
 \bra{\psi^L_{\mathscr{r}\mu\sigma}}
 \mathcal K_{\rm F}(q_{\mathscr{r}\mu\sigma};f_\mathscr{r})
 &=
 \omega\bra{\psi^L_{\mathscr{r}\mu\sigma}}.
 \label{eq:app_lead_eigenproblem_v9}
\end{align}
The superscripts $R$ and $L$ denote right and left eigenvectors only.
They do not denote propagation direction or a spatial side of the
interface.  The spinors are chosen biorthogonally,
\begin{equation}
 \braket{
 \psi^L_{\mathscr{r}\mu\sigma}
 |
 \psi^R_{\mathscr{r}\nu\sigma'}
 }
 =
 \delta_{\mu\nu}\delta_{\sigma\sigma'}.
 \label{eq:app_lead_biorth_v9}
\end{equation}
Writing the right eigenvector as
\begin{equation}
 \ket{\psi^R_{\mathscr{r}\mu\sigma}}
 =
 \begin{pmatrix}
 \bm u_{\mathscr{r}\mu\sigma}\\
 \bm v_{\mathscr{r}\mu\sigma}
 \end{pmatrix}
 =
 \begin{pmatrix}
 u_{\mathscr{r}\mu\sigma,A}\\
 u_{\mathscr{r}\mu\sigma,B}\\
 v_{\mathscr{r}\mu\sigma,A}\\
 v_{\mathscr{r}\mu\sigma,B}
 \end{pmatrix},
 \label{eq:app_mode_spinor_v9}
\end{equation}
the biorthogonal flux velocity is
\begin{align}
 v_{\mathscr{r}\mu\sigma}(\omega)
 &=
 \Re\!\left[
 \bra{\psi^L_{\mathscr{r}\mu\sigma}}
 \partial_q\mathcal K_{\rm F}(q;f_\mathscr{r})
 \ket{\psi^R_{\mathscr{r}\mu\sigma}}
 \right]_{q=q_{\mathscr{r}\mu\sigma}},
 \label{eq:app_flux_velocity_consistent}
 \\
 \partial_q\mathcal K_{\rm F}(q;f_\mathscr{r})
 &=
 iT_+(f_\mathscr{r})e^{iq}
 -iT_-(f_\mathscr{r})e^{-iq}.
 \nonumber
\end{align}
The sign of $v_{\mathscr{r}\mu\sigma}$ determines the propagation direction.

The labels $\sigma=+$ and $\sigma=-$ denote the particle-like and
hole-like Nambu sectors, respectively. Numerically, this
classification is obtained from the particle-hole polarization of
the Nambu spinor. The sector label and the propagation label are
independent: the incoming or outgoing character of a propagating root
is assigned only after its flux has been evaluated.

Before assembling the scattering basis, the right eigenvectors are
rescaled by $|v_{\mathscr{r}\mu\sigma}|^{-1/2}$.  The corresponding dual
left eigenvectors are scaled inversely, so that their biorthogonal
overlap remains unity.  We denote these flux-normalized vectors by
$\ket{\widetilde\psi^R_{\mathscr{r}\mu\sigma}}$ and
$\bra{\widetilde\psi^L_{\mathscr{r}\mu\sigma}}$.

\section{Channel operators and their relation to the lattice field}
\label{app:channel_operators_v9}

The asymptotic channel operators are coefficients in the mode
expansion of the same physical lattice field introduced in the main
text; they are not additional microscopic degrees of freedom.  In a
uniform asymptotic region $\mathscr{r}$, the Fourier component of the
physical annihilation spinor reads
\begin{equation}
 \hat{\bm\Psi}_\mathscr{r}(q)
 =
 \frac{1}{\sqrt{N_\mathscr{r}^{\rm as}}}
 \sum_{j\in\mathscr{r}}
 e^{-iq(j-j_\mathscr{r})}
 \hat{\bm\Psi}_j,
 \label{eq:app_lead_Fourier_Psi}
\end{equation}
where $j_\mathscr{r}$ is a reference cell and $N^{as}_\mathscr{r}$ is the number of cells used to define the asymptotic Fourier transform.  The corresponding Nambu operator is
\begin{equation}
 \hat{\bm\Upsilon}_\mathscr{r}(q)
 =
 \left(
 \hat a_{\mathscr{r},q},
 \hat b_{\mathscr{r},q},
 \hat a_{\mathscr{r},-q}^\dagger,
 \hat b_{\mathscr{r},-q}^\dagger
 \right)^{T}.
 \label{eq:app_lead_Fourier_Upsilon}
\end{equation}
The first two components of $\hat{\bm\Upsilon}_\mathscr{r}(q)$ are precisely
the components of $\hat{\bm\Psi}_\mathscr{r}(q)$; the last two form the
corresponding creation sector at momentum $-q$.

For a particle-like root, the channel annihilation
operator is obtained by projection onto the corresponding
flux-normalized dual mode,
\begin{equation}
 \hat\eta_{\mathscr{r}\mu,+}(\omega)
 =
 \bra{\widetilde\psi^L_{\mathscr{r}\mu,+}}
 \hat{\bm\Upsilon}_\mathscr{r}
 \!\left(q_{\mathscr{r}\mu,+},\omega\right).
 \label{eq:app_eta_plus_projection_v9}
\end{equation}
For a hole-like root, the coefficient appearing in
the Nambu expansion is a creation operator,
\begin{equation}
 \hat\eta_{\mathscr{r}\mu,-}^\dagger(\omega)
 =
 \bra{\widetilde\psi^L_{\mathscr{r}\mu,-}}
 \hat{\bm\Upsilon}_\mathscr{r}
 \!\left(q_{\mathscr{r}\mu,-},\omega\right).
 \label{eq:app_eta_minus_projection_v9}
\end{equation}
These equations make explicit the relation between the channel
operators and the microscopic lattice operators
$\hat a_j,\hat b_j,\hat a_j^\dagger,\hat b_j^\dagger$.

Restricting to propagating asymptotic channels, the Nambu
field operator can therefore be expanded as
\begin{align}
 \hat{\bm\Upsilon}_\mathscr{r}(j,t)
 &=
 \int_0^\infty\frac{d\omega}{\sqrt{2\pi}}\,
 e^{-i\omega t}
 \Bigg[
 \sum_{\mu}
 \widetilde{\bm\psi}^R_{\mathscr{r}\mu,+}(\omega)
 e^{iq_{\mathscr{r}\mu,+}(\omega)(j-j_\mathscr{r})}
 \hat\eta_{\mathscr{r}\mu,+}(\omega)
 \nonumber\\
 &\qquad\qquad
 +
 \sum_{\mu}
 \widetilde{\bm\psi}^R_{\mathscr{r}\mu,-}(\omega)
 e^{iq_{\mathscr{r}\mu,-}(\omega)(j-j_\mathscr{r})}
 \hat\eta_{\mathscr{r}\mu,-}^\dagger(\omega)
 \Bigg]
 +\text{h.c.-related frequency sector}.
 \label{eq:app_asymptotic_field_expansion_v9}
\end{align}
Evanescent roots are retained in the interface matching because they
are required to satisfy the boundary equations, but they do not
represent independent asymptotic input or output operators and are
not included as propagating channels in the scattering matrix.

\section{Incoming and outgoing channels and the scattering matrix}
\label{app:interface_scattering_v8}

At each real frequency, all propagating roots are first classified by
their Nambu sector and then by the direction of their biorthogonal
flux.  Coefficients of modes propagating toward the inhomogeneous
region are collected into the incoming operator vector, while those
propagating away from it are collected into the outgoing vector,
\begin{equation}
 \hat{\bm A}_{\rm in}(\omega)
 =
 \begin{pmatrix}
 \hat{\bm\eta}_{+,{\rm in}}(\omega)\\
 \hat{\bm\eta}_{-,{\rm in}}^\dagger(\omega)
 \end{pmatrix},
 \qquad
 \hat{\bm A}_{\rm out}(\omega)
 =
 \begin{pmatrix}
 \hat{\bm\eta}_{+,{\rm out}}(\omega)\\
 \hat{\bm\eta}_{-,{\rm out}}^\dagger(\omega)
 \end{pmatrix}.
 \label{eq:app_Ain_Aout_v9}
\end{equation}
The fixed-frequency scattering matrix is defined by
\begin{equation}
 \hat{\bm A}_{\rm out}(\omega)
 =
 \mathcal S(\omega)\hat{\bm A}_{\rm in}(\omega),
 \qquad
 \mathcal S(\omega)
 =
 \begin{pmatrix}
 \mathcal S_{++}&\mathcal S_{+-}\\
 \mathcal S_{-+}&\mathcal S_{--}
 \end{pmatrix}.
 \label{eq:app_S_v8}
\end{equation}
The block notation is
$\mathcal S_{\sigma_{\rm out}\sigma_{\rm in}}$: the first index
specifies the Nambu sector of the outgoing channel and the second the
sector of the incoming channel.  Hence, in the fermionic problem,
$\mathcal S_{+-}$ maps an incoming hole-like channel to an outgoing
particle-like channel, while $\mathcal S_{-+}$ describes the reverse
sector-changing process.  The diagonal blocks describe scattering
within the same Nambu sector.

If $\mathsf P_\sigma^{\rm in}$ and
$\mathsf P_\sigma^{\rm out}$ select a sector $\sigma$ from the complete incoming and outgoing channel vectors, respectively, then
\begin{equation}
 \mathcal S_{\sigma_{\rm out}\sigma_{\rm in}}
 =
 \mathsf P_{\sigma_{\rm out}}^{\rm out}
 \mathcal S
 \mathsf P_{\sigma_{\rm in}}^{{\rm in}\, T}.
 \label{eq:app_S_blocks_projectors_v9}
\end{equation}
Equivalently,
\begin{align}
 \hat\eta_{\mu,+,{\rm out}}
 &=
 \sum_{\nu}
 (\mathcal S_{++})_{\mu\nu}
 \hat\eta_{\nu,+,{\rm in}}
 +
 \sum_{\nu}
 (\mathcal S_{+-})_{\mu\nu}
 \hat\eta_{\nu,-,{\rm in}}^\dagger,
 \label{eq:app_S_first_row_v9}
 \\
 \hat\eta_{\mu,-,{\rm out}}^\dagger
 &=
 \sum_{\nu}
 (\mathcal S_{-+})_{\mu\nu}
 \hat\eta_{\nu,+,{\rm in}}
 +
 \sum_{\nu}
 (\mathcal S_{--})_{\mu\nu}
 \hat\eta_{\nu,-,{\rm in}}^\dagger.
 \label{eq:app_S_second_row_v9}
\end{align}

\section{Finite-interface matching}
\label{app:finite_matching_consistent}

For the numerical construction of $\mathcal S(\omega)$, we now work
with c-number channel amplitudes rather than operators.  This
distinction avoids confusing the operator vectors
$\hat{\bm A}_{\rm in/out}$ with the numerical boundary coefficients.
Let $L_{\rm d}$ be the number of cells retained in the finite
inhomogeneous region and define
\begin{equation}
 \bm X(\omega)
 =
 \begin{pmatrix}
 \bm\zeta_1\\
 \vdots\\
 \bm\zeta_{L_{\rm d}}\\
 \bm a_{\rm out}^{(\mathrm{left})}\\
 \bm a_{\rm out}^{(\mathrm{right})}
 \end{pmatrix}.
 \label{eq:app_unknown_vector_consistent}
\end{equation}
The two boundary-amplitude vectors include both propagating outgoing
modes and the decaying evanescent modes required at the corresponding
boundary.  The block-tridiagonal interface equations and the two
boundary conditions can be written as
\begin{equation}
 \mathbb M(\omega)\bm X(\omega)
 =
 \mathbb B_{\rm in}(\omega)\bm a_{\rm in}(\omega),
 \label{eq:app_linear_matching_system_v9}
\end{equation}
where $\bm a_{\rm in}$ contains the prescribed incoming propagating
amplitudes.  The matrix $\mathbb M$ contains the real-space blocks
$V_{\rm F}$, $T_+$, and $T_-$ together with the outgoing and evanescent
boundary modes, while $\mathbb B_{\rm in}$ injects the chosen incoming
waves.

If $\mathbb P_{\rm out}$ selects the outgoing boundary coefficients
from $\bm X$ and $\mathbb P_{\rm prop}$ removes the evanescent
components, then
\begin{equation}
 \mathcal S(\omega)
 =
 \mathbb P_{\rm prop}
 \mathbb P_{\rm out}
 \mathbb M^{-1}(\omega)
 \mathbb B_{\rm in}(\omega).
 \label{eq:app_S_from_matching_v9}
\end{equation}
The inverse is not formed explicitly in the numerical implementation.
Instead, $\mathbb M(\omega)$ is factorized once at each frequency and
the linear system is solved separately for each unit incoming
amplitude.  The resulting columns are then sorted by Nambu sector and
propagation direction to form the blocks of
Eq.~\eqref{eq:app_S_v8}.

\section{Incoming vacuum and selected Hawking-partner moments}
\label{app:incoming_vacuum_consistent}

The incoming channel vacuum is defined by
\begin{align}
 \left\langle
 \hat\eta_{a,{\rm in}}^\dagger(\omega)
 \hat\eta_{b,{\rm in}}(\omega')
 \right\rangle
 &=0,
 \nonumber\\
 \left\langle
 \hat\eta_{a,{\rm in}}(\omega)
 \hat\eta_{b,{\rm in}}^\dagger(\omega')
 \right\rangle
 &=
 2\pi\delta_{ab}\delta(\omega-\omega').
 \label{eq:app_incoming_vacuum_v9}
\end{align}
Suppressing the common frequency delta function, the scattering
relations give
\begin{align}
 \left\langle
 \hat\eta_{\nu,+,{\rm out}}^\dagger
 \hat\eta_{\mu,+,{\rm out}}
 \right\rangle
 &=
 \left[
 \mathcal S_{+-}\mathcal S_{+-}^\dagger
 \right]_{\mu\nu},
 \label{eq:app_nplus_matrix_v9}
 \\
 \left\langle
 \hat\eta_{\nu,-,{\rm out}}^\dagger
 \hat\eta_{\mu,-,{\rm out}}
 \right\rangle
 &=
 \left[
 \mathcal S_{-+}\mathcal S_{-+}^\dagger
 \right]_{\mu\nu},
 \label{eq:app_nminus_matrix_v9}
 \\
 \left\langle
 \hat\eta_{\mu,+,{\rm out}}
 \hat\eta_{\nu,-,{\rm out}}
 \right\rangle
 &=
 \left[
 \mathcal S_{++}\mathcal S_{-+}^\dagger
 \right]_{\mu\nu}.
 \label{eq:app_mplusminus_matrix_v9}
\end{align}

Consistently with Eq.~\eqref{eq:secIV_selected_eta_F}, the selected
outgoing Hawking and partner channels are
\begin{equation}
 \hat\eta_H(\omega)
 =
 \hat\eta_{\mathscr{r}_H\mu_H+,{\rm out}}(\omega),
 \qquad
 \hat\eta_P(\omega)
 =
 \hat\eta_{\mathscr{r}_P\mu_P-,{\rm out}}(\omega).
 \label{eq:app_selected_channel_operators_consistent}
\end{equation}
Their normal occupations and anomalous pair moment are
\begin{align}
 n_H(\omega)
 &=
 \left[
 \mathcal S_{+-}(\omega)\mathcal S_{+-}^\dagger(\omega)
 \right]_{HH},
 \nonumber\\
 n_P(\omega)
 &=
 \left[
 \mathcal S_{-+}(\omega)\mathcal S_{-+}^\dagger(\omega)
 \right]_{PP},
 \nonumber\\
 m_{HP}(\omega)
 &=
 \left[
 \mathcal S_{++}(\omega)\mathcal S_{-+}^\dagger(\omega)
 \right]_{HP}
 =
 \left\langle
 \hat\eta_H(\omega)\hat\eta_P(\omega)
 \right\rangle.
 \label{eq:app_channel_moments_v8}
\end{align}
For the conditional incoming vacuum, the quantity plotted as the
selected Hawking-channel conversion is simply $n_H(\omega)$. Thus, $n_H$ is one selected diagonal matrix element and not
the trace of the full sector-changing block.

\section{Real-space reconstruction and fermionic Wick contraction}
\label{app:covariance_wick_v8}

The selected channel operators are related back to the physical
lattice basis through their annihilation-sector envelopes.  Around
the first horizon,
\begin{align}
 \hat\Psi_{(n,s)}(\omega)
 &\supset
 \varphi_{H,s}(n,\omega)\hat\eta_H(\omega),
 \qquad n>r_1,
 \nonumber\\
 \hat\Psi_{(m,s)}(\omega)
 &\supset
 \varphi_{P,s}(m,\omega)\hat\eta_P(\omega),
 \qquad m<r_1.
 \label{eq:app_selected_realspace_expansion_v9}
\end{align}
For the selected right eigenmodes, write
\begin{equation}
 \ket{\psi^R_{\alpha}}
 =
 \left(
 u_{\alpha,A},
 u_{\alpha,B},
 v_{\alpha,A},
 v_{\alpha,B}
 \right)^{T},
 \qquad
 \alpha\in\{H,P\}.
 \label{eq:app_selected_spinors_consistent}
\end{equation}
The envelopes used in the reconstruction are
\begin{align}
 \varphi_{H,s}(n,\omega)
 &=
 u_{H,s}(\omega)
 e^{iq_H(\omega)(n-r_1)},
 \label{eq:app_phiH_consistent}
 \\
 \varphi_{P,s}(m,\omega)
 &=
 v_{P,s}^*(\omega)
 e^{-iq_P(\omega)(m-r_1)},
 \label{eq:app_phiP_consistent}
\end{align}
with $s\in\{A,B\}$.  The conjugation of the hole component in
Eq.~\eqref{eq:app_phiP_consistent} converts the selected hole-like
Nambu mode into the annihilation representation used in the lattice
covariance.

The anomalous Hawking-partner contribution is therefore
\begin{equation}
 F^{HP}_{ms,ns'}
 =
 \int_{\omega_a}^{\omega_b}
 \frac{d\omega}{2\pi}\,
 w(\omega)m_{HP}(\omega)
 \varphi_{P,s}(m,\omega)
 \varphi_{H,s'}(n,\omega),
 \label{eq:app_F_reconstruction_consistent}
\end{equation}
where $w(\omega)$ is the smooth sine-squared endpoint window used
numerically to suppress finite-window ringing.  Equation
\eqref{eq:app_F_reconstruction_consistent} is the detailed form of
Eq.~\eqref{eq:secIV_F_reconstruction_F}.

Using the same lattice notation as in the main text,
\begin{equation}
 \Gamma_{pq}
 =
 \langle\hat\Psi_q^\dagger\hat\Psi_p\rangle,
 \qquad
 F_{pq}
 =
 \langle\hat\Psi_p\hat\Psi_q\rangle,
 \qquad
 \bar F_{pq}
 =
 \langle\hat\Psi_p^\dagger\hat\Psi_q^\dagger\rangle.
 \label{eq:app_covariances_consistent}
\end{equation}
For fermions,
\begin{equation}
 F^{T}=-F,
 \qquad
 \bar F=F^\dagger.
 \label{eq:app_F_fermion_symmetry}
\end{equation}
For two distinct lattice modes $p\neq q$, Wick's theorem gives
\begin{align}
 G_{pq,{\rm conn}}^{(2),{\rm F}}
 &=
 -\Gamma_{pq}\Gamma_{qp}
 -\bar F_{pq}F_{pq}.
 \label{eq:app_WickF_intermediate}
\end{align}
Since $\Gamma_{qp}=\Gamma_{pq}^*$ and
$\bar F_{pq}=F_{qp}^*=-F_{pq}^*$, this becomes
\begin{equation}
 G_{pq,{\rm conn}}^{(2),{\rm F}}
 =
 -|\Gamma_{pq}|^2
 +|F_{pq}|^2,
 \label{eq:app_WickF_final}
\end{equation}
which reproduces Eq.~\eqref{eq:secIV_WickF_only}.

The numerical reconstruction uses $D=48$ cells on each side of each
interface. Around the first horizon at $r_1=60$, the partner window is
$12\leq m\leq59$ and the Hawking window is
$61\leq n\leq108$. In the embedded fermionic map of
Fig.~\ref{fig:secIV_cross_F}(a), these finite reconstructed blocks are
placed at their physical positions inside an $N\times N$ array, while
all unreconstructed entries are filled with zero. The abrupt outer
edges of the colored blocks therefore mark the finite reconstruction
window and are not physical discontinuities of a global covariance.

\begin{figure*}[t]
 \centering
 \includegraphics[width=0.92\textwidth]
 {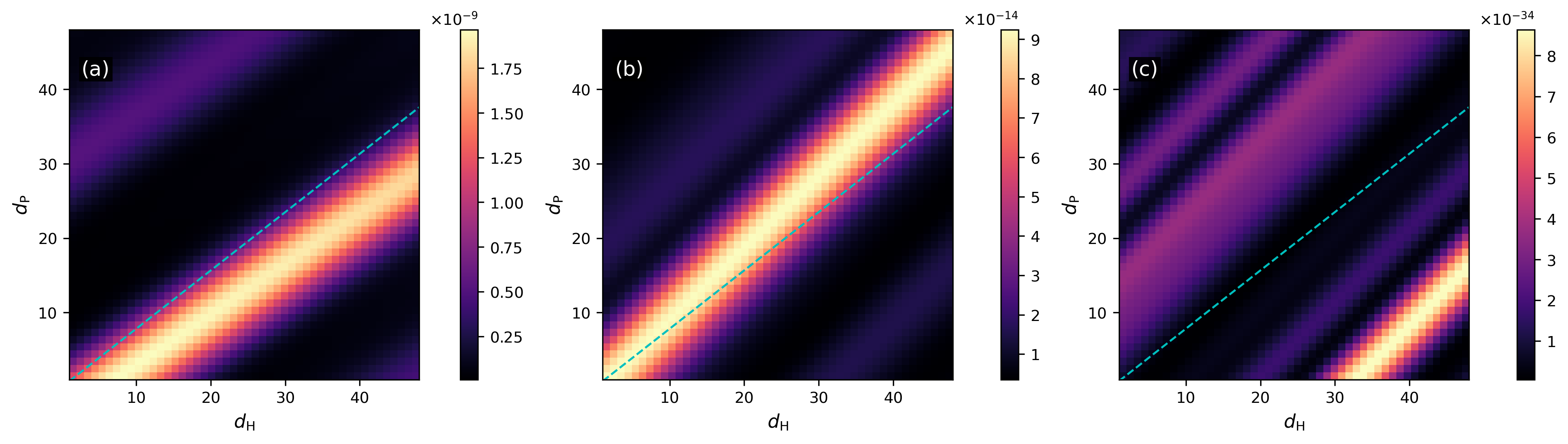}
 \caption{Isolated fermionic cross-interface blocks used in the real-space
 reconstruction. (a)-(c) The horizon, no-crossing, and
 uniform profiles, respectively. This representation removes the
 zero-filled unreconstructed parts of the embedded $N\times N$ map
 and makes the finite computational support explicit. The cyan dashed
 line follows the independently calculated equal-travel-time relation,
 and each panel has its own colorbar.}
 \label{fig:app_cross_only_v8}
\end{figure*}

\section{Fermionic covariance-positivity constraint}
\label{app:frequency_window_v8}

The bound used in Sec.~\ref{subsec:fermionic_physicality} follows
directly from positivity and the fermionic canonical
anticommutation relations.  Applying the operator Cauchy-Schwarz
inequality to
$A=\hat\eta_H^\dagger$ and $B=\hat\eta_P$ gives
\begin{equation}
 |m_{HP}|^2
 \leq
 \langle\hat\eta_H\hat\eta_H^\dagger\rangle
 \langle\hat\eta_P^\dagger\hat\eta_P\rangle
 =
 (1-n_H)n_P.
 \label{eq:app_Pauli_bound_1}
\end{equation}
Interchanging the two channels gives
\begin{equation}
 |m_{HP}|^2
 \leq
 n_H(1-n_P).
 \label{eq:app_Pauli_bound_2}
\end{equation}
Both inequalities must hold simultaneously, hence
\begin{equation}
 |m_{HP}|^2
 \leq
 \min\!\left[
 n_H(1-n_P),
 (1-n_H)n_P
 \right].
 \label{eq:app_fermionic_Pauli_v10}
\end{equation}
This is the bound used to define $\Theta_{HP}^{\rm F}$ in the main
text. It is a fermionic covariance-positivity condition rather than a nonclassicality witness. The general covariance-matrix formulation of
fermionic Gaussian states is discussed, for example, in
Ref.~\cite{Bravyi2005}. A genuinely quantum fermionic Hawking claim
requires an independent entanglement criterion. One suitable choice
is a fermionic logarithmic negativity based on a fermionic partial
transpose or partial-time-reversal construction
\cite{ShapourianShiozakiRyu2017,ShapourianRyu2019}; this quantity is
not evaluated in the present work.

The horizon calculation uses the profile defined in
Eq.~\eqref{eq:horizonprofile}, with $\kappa_1=1.2$ and
$\kappa_2=0.8$.  The no-crossing control uses the same profile,
interface positions, and width but sets
$\kappa_1^{\rm nc}=0.98$ and $\kappa_2^{\rm nc}=0.94$, so that the
critical value is not crossed.  The uniform control sets
$f(j)=1.2$ for every cell.  The pairing amplitude, chain length,
frequency grid, and channel-selection algorithm are otherwise kept
unchanged.

The scattering problem was first scanned over
$0.005\leq\omega/\tau\leq2.15$.  The interval
$0.06\leq\omega/\tau\leq0.95$ was then selected because the chosen
Hawking and partner roots $q_H(\omega)$ and $q_P(\omega)$ remain on
the first smooth low-energy propagating solution pair throughout this
window.  Frequencies very close to zero are excluded because the
flux normalization becomes sensitive to small velocities, while the
higher-frequency region is excluded because the ordering and number
of propagating roots change.  This interval is an analysis window and
is not interpreted as an analogue Hawking temperature.

\begin{figure*}[t]
 \centering
 \includegraphics[width=0.82\textwidth]
 {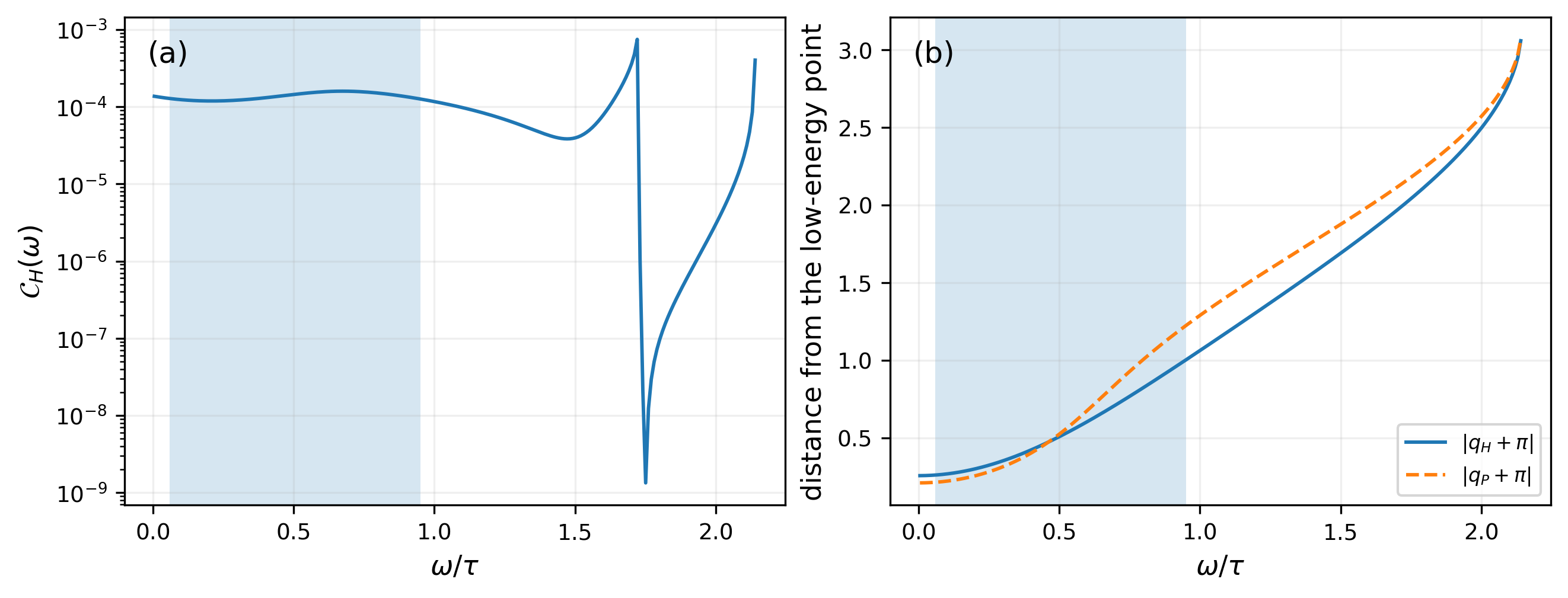}
 \caption{Broad fermionic frequency-window diagnostic. (a) The selected Hawking-channel quantity $\mathcal C_H=n_H$ over the
 broad scan; the shaded region denotes the interval retained in the
 main calculation. (b) The distance of the selected
 Hawking and partner wave numbers from the low-energy point $q=-\pi$.
 The window $0.06\leq\omega/\tau\leq0.95$ follows the first smooth
 pair of selected propagating roots and avoids both the near-zero
 normalization-sensitive region and the higher-frequency channel
 rearrangements.}
 \label{fig:app_frequency_window_v8}
\end{figure*}

A narrower interval,
$0.70\leq\omega/\tau\leq0.90$, is used only as a kinematic
diagnostic.  Across the broader interval, the ratio
$|v_P(\omega)/v_H(\omega)|$ changes with frequency, so the integrated
real-space covariance contains a superposition of nearby ridge
slopes.  Restricting the interval reduces this superposition and
tests whether the remaining ridge follows the independently
calculated equal-travel-time relation.

\begin{figure*}[t]
 \centering
 \includegraphics[width=0.96\textwidth]
 {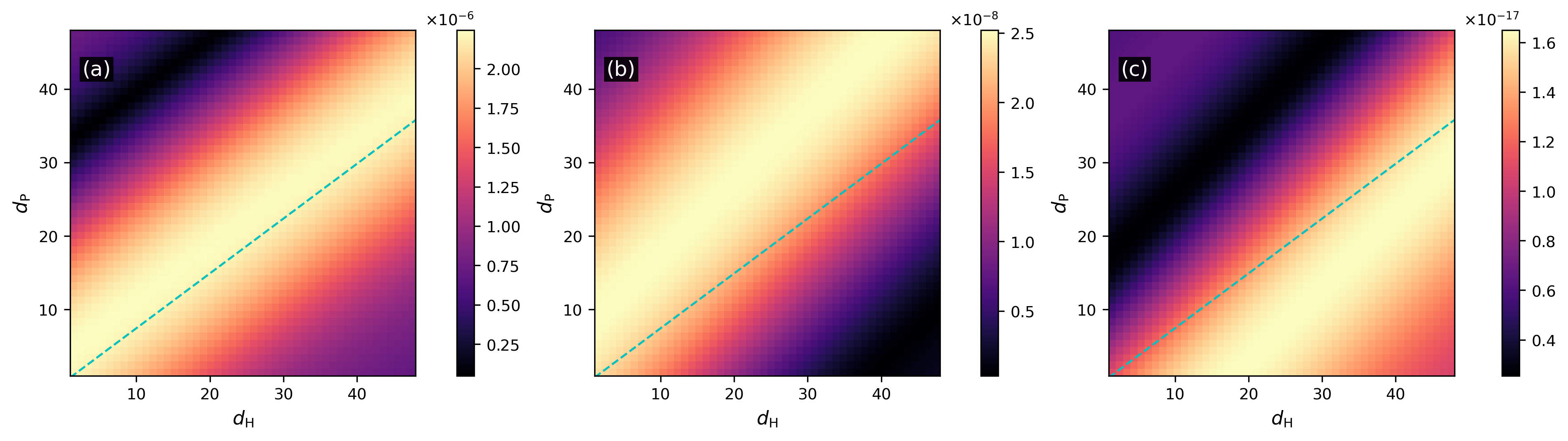}
 \caption{Fermionic narrowband kinematic check over
 $0.70\leq\omega/\tau\leq0.90$. Panels (a)-(c) show the horizon,
 no-crossing, and uniform profiles, respectively. The plotted quantity
 is $[\sum_{ss'}|F_{ms,ns'}|^2]^{1/2}$. The cyan dashed line follows
 the velocity ratio obtained independently from the selected
 asymptotic channels. This figure tests the propagation geometry only
 and is not used to define the selected frequency interval.}
 \label{fig:app_narrowband_v8}
\end{figure*}

For the conditional incoming vacuum,
$\mathcal C_H^X(\omega)=n_H^X(\omega)$, so the integrated quantity
used in the main text is
\begin{equation}
 \mathcal I_H^X
 =
 \int_{\omega_a}^{\omega_b}
 d\omega\,n_H^X(\omega),
 \qquad
 X\in
 \{
 {\rm horizon},
 {\rm no\text{-}crossing},
 {\rm uniform}
 \}.
 \label{eq:app_integrated_nH_consistent}
\end{equation}
The enhancement factor compares the horizon profile with the
no-crossing interface, while retaining the same numerical and pairing
parameters. It therefore quantifies horizon selectivity of the
selected channel, but not thermality or nonclassicality.

\section{Canonical benchmark}
\label{app:canonical_completion_v8}

The scattering matrix calculated above belongs to the conditional
non-Hermitian problem and is not assumed to preserve the canonical
output algebra. For fermions, a canonical Bogoliubov transformation
must preserve the canonical anticommutation relations and is unitary
in the extended channel basis. As a sensitivity benchmark, the code
computes the unitary polar factor $U$ of the conditional matrix,
\begin{equation}
 \mathcal S=U P_{\rm pol},
 \qquad
 P_{\rm pol}=(\mathcal S^\dagger\mathcal S)^{1/2}.
 \label{eq:app_polar_decomposition_consistent}
\end{equation}
For a nonsingular matrix, $U$ is the nearest unitary matrix in the
Frobenius norm \cite{Higham1986}. We therefore refer to this
construction as a \emph{unitary polar benchmark}; it is not a
microscopic reservoir completion. The horizon-selective spatial
correlations and their suppression in the control profiles do not
rely on this benchmark, which is used only to test the sensitivity of
the fermionic channel ratios to imposing a canonical unitary
constraint.
\end{appendix}
\bibliography{HRNHMA}

@BOOK{BirrelDavies-QFCST,
author = "N.D. Birrel and P.C.W. Davies",
title = "Quantum Fields in Curved Space chap 2",
volume = "1ed",
publisher = "Cambridge University Press",
year = "1984"}

@Article{HelbigHofman2020,
author={Helbig, T.
and Hofmann, T.
and Imhof, S.
and Abdelghany, M.
and Kiessling, T.
and Molenkamp, L. W.
and Lee, C. H.
and Szameit, A.
and Greiter, M.
and Thomale, R.},
title={Generalized bulk-boundary correspondence in non-{Hermitian} topolectrical circuits},
journal={Nat. Phys.},
year={2020},
month={Jul},
day={01},
volume={16},
number={7},
pages={747-750},
issn={1745-2481},
doi={10.1038/s41567-020-0922-9},
url={https://doi.org/10.1038/s41567-020-0922-9}
}

@article{SlootmanCherifi2024,
  title = {Breaking and resurgence of symmetry in the non-{Hermitian} {Su}-{Schrieffer}-{Heeger} model in photonic waveguides},
  author = {Slootman, E. and Cherifi, W. and Eek, L. and Arouca, R. and Bergholtz, E. J. and Bourennane, M. and Smith, C. Morais},
  journal = {Phys. Rev. Res.},
  volume = {6},
  issue = {2},
  pages = {023140},
  numpages = {23},
  year = {2024},
  month = {May},
  publisher = {American Physical Society},
  doi = {10.1103/PhysRevResearch.6.023140},
  url = {https://link.aps.org/doi/10.1103/PhysRevResearch.6.023140}
}

@book{wald1994QFTCST,
  title = {Quantum Field Theory in Curved Spacetime and Black Hole Thermodynamics},
  author = {Wald, Robert M.},
  year = {1994},
  publisher = {University of Chicago Press},
  address = {Chicago},
  isbn = {9780226870274},
  url = {https://press.uchicago.edu/ucp/books/book/chicago/Q/bo3684008.html}
}

@article{hawking_particle_1975,
    title = {Particle creation by black holes},
    volume = {43},
    issn = {1432-0916},
    url = {https://doi.org/10.1007/BF02345020},
    doi = {10.1007/BF02345020},
    number = {3},
    urldate = {2025-06-01},
    journal = {Commun. Math. Phys.},
    author = {Hawking, S. W.},
    month = aug,
    year = {1975},
    pages = {199--220},
}

@article{barcelo_analogue_2011,
    title = {Analogue {Gravity}},
    volume = {14},
    issn = {1433-8351},
    url = {https://doi.org/10.12942/lrr-2011-3},
    doi = {10.12942/lrr-2011-3},
    number = {1},
    urldate = {2025-06-02},
    journal = {Living Rev. Relativ.},
    author = {Barceló, Carlos and Liberati, Stefano and Visser, Matt},
    month = may,
    year = {2011},
    pages = {3},
}

@article{unruh_experimental_1981,
    title = {Experimental {Black}-{Hole} {Evaporation}?},
    volume = {46},
    copyright = {http://link.aps.org/licenses/aps-default-license},
    issn = {0031-9007},
    url = {https://link.aps.org/doi/10.1103/PhysRevLett.46.1351},
    doi = {10.1103/PhysRevLett.46.1351},
    number = {21},
    urldate = {2025-05-26},
    journal = {Phys. Rev. Lett.},
    author = {Unruh, W. G.},
    month = may,
    year = {1981},
    pages = {1351--1353},
}

@article{steinhauer_observation_2016,
    title = {Observation of quantum {Hawking} radiation and its entanglement in an analogue black hole},
    volume = {12},
    copyright = {2016 Springer Nature Limited},
    issn = {1745-2481},
    url = {https://www.nature.com/articles/nphys3863},
    doi = {10.1038/nphys3863},
    number = {10},
    urldate = {2025-05-26},
    journal = {Nat. Phys.},
    author = {Steinhauer, Jeff},
    month = oct,
    year = {2016},
    pages = {959--965},
}

@ARTICLE{Stalhammar-NJP2023,
        title = "$\mathcal{PT}$ symmetry-protected exceptional cones and analogue {Hawking} radiation",
        author = "St\aa{}lhammar, Marcus and Larana-Aragon, Jorge and Rødland, Lukas and Kunst, Flore K",
        journal = "New J. Phys.",
        volume = "25",
        number = "4",
        pages ="043012",
        year = "2023",
        month = "apr",
        publisher = "IOP Publishing",
        doi = "10.1088/1367-2630/acc6e5",
        url = "https://dx.doi.org/10.1088/1367-2630/acc6e5"}

@article{BalbinotEtAl2008,
  author  = {Balbinot, Roberto and Fabbri, Alessandro and Fagnocchi, Serena and Recati, Alessio and Carusotto, Iacopo},
  title   = {Nonlocal density correlations as a signature of {Hawking} radiation from acoustic black holes},
  journal = {Phys. Rev. A},
  year    = {2008},
  volume  = {78},
  pages   = {021603},
  doi     = {10.1103/PhysRevA.78.021603}
}

@article{CarusottoEtAl2008NJP,
  author  = {Carusotto, Iacopo and Fagnocchi, Serena and Recati, Alessio and Balbinot, Roberto and Fabbri, Alessandro},
  title   = {Numerical observation of {Hawking} radiation from acoustic black holes in atomic {Bose}-{Einstein} condensates},
  journal = {New J. Phys.},
  year    = {2008},
  volume  = {10},
  pages   = {103001},
  doi     = {10.1088/1367-2630/10/10/103001}
}

@article{MacherParentani2009,
  author  = {Macher, Jean and Parentani, Renaud},
  title   = {Black-hole radiation in {Bose}-{Einstein} condensates},
  journal = {Phys. Rev. A},
  year    = {2009},
  volume  = {80},
  pages   = {043601},
  doi     = {10.1103/PhysRevA.80.043601}
}

@article{RecatiPavloffCarusotto2009,
  author       = {Recati, A. and Pavloff, N. and Carusotto, I.},
  title        = {Bogoliubov theory of acoustic {H}awking radiation in {B}ose--{E}instein condensates},
  journal      = {Phys. Rev. A},
  volume       = {80},
  pages        = {043603},
  year         = {2009},
  doi          = {10.1103/PhysRevA.80.043603},
  
}

@article{deNova2019,
  author  = {Mu{\~n}oz de Nova, Juan Ram{\'o}n and Golubkov, Katrine and Kolobov, Victor I. and Steinhauer, Jeff},
  title   = {Observation of thermal {Hawking} radiation and its temperature in an analogue black hole},
  journal = {Nature},
  year    = {2019},
  volume  = {569},
  pages   = {688--691},
  doi     = {10.1038/s41586-019-1241-0}
}

@article{Unruh1976,
  author       = {Unruh, W. G.},
  title        = {Notes on black-hole evaporation},
  journal      = {Phys. Rev. D},
  year         = {1976},
  volume       = {14},
  pages        = {870--892},
  doi          = {10.1103/PhysRevD.14.870}
}

@article{Visser1998,
  author  = {Visser, Matt},
  title   = {Acoustic black holes: Horizons, ergospheres and {Hawking} radiation},
  journal = {Class. Quantum Gravity},
  year    = {1998},
  volume  = {15},
  pages   = {1767--1791},
  doi     = {10.1088/0264-9381/15/6/024}
}

@book{FabbriNavarro2005,
  author    = {Fabbri, Alessandro and Navarro-Salas, Jos{\'e}},
  title     = {Modeling Black Hole Evaporation},
  publisher = {Imperial College Press},
  address   = {London},
  year      = {2005},
  isbn      = {978-1848160476},
  doi       = {10.1142/p378},
  url       = {https://doi.org/10.1142/p378}
}

@article{Steinhauer2015_PRD,
  author       = {Steinhauer, Jeff},
  title        = {Measuring the entanglement of analogue {Hawking} radiation
                  by the density-density correlation function},
  journal      = {Phys. Rev. D},
  volume       = {92},
  number       = {2},
  pages        = {024043},
  year         = {2015},
  doi          = {10.1103/PhysRevD.92.024043}
}

@article{MunozArboleda2026ThermodynamicsAnalogue,
  author  = {Munoz-Arboleda, D. F. and St{\aa}lhammar, M. and Morais Smith, C.},
  title   = {Thermodynamics of analogue black holes in a non-Hermitian tight-binding model},
  journal = {Phys. Rev. B},
  volume  = {113},
  pages   = {L081110},
  year    = {2026},
  doi     = {10.1103/vdsx-r3dq},
  url     = {https://doi.org/10.1103/vdsx-r3dq}
}

@article{brody2014biorthogonal,
  author  = {Brody, Dorje C.},
  title   = {Biorthogonal quantum mechanics},
  journal = {J. Phys. A: Math. Theor.},
  volume  = {47},
  number  = {3},
  pages   = {035305},
  year    = {2014},
  doi     = {10.1088/1751-8113/47/3/035305},
  url     = {https://doi.org/10.1088/1751-8113/47/3/035305}
}

@article{GarayEtAl2000,
  author  = {Garay, L. J. and Anglin, J. R. and Cirac, J. I. and Zoller, P.},
  title   = {Sonic Analog of Gravitational Black Holes in {Bose}-{Einstein} Condensates},
  journal = {Phys. Rev. Lett.},
  volume  = {85},
  pages   = {4643--4647},
  year    = {2000},
 doi      = {10.1103/PhysRevLett.85.4643},
  url     = {https://doi.org/10.1103/PhysRevLett.85.4643},  
}

@article{LahavEtAl2010,
  author  = {Lahav, Oren and Itah, Amir and Blumkin, Alex and Gordon, Carmit and Rinott, Shahar and Zayats, Alona and Steinhauer, Jeff},
  title   = {Realization of a Sonic Black Hole Analog in a {Bose}-{Einstein} Condensate},
  journal = {Phys. Rev. Lett.},
  volume  = {105},
  pages   = {240401},
  year    = {2010},
  doi     = {10.1103/PhysRevLett.105.240401},
  url     = {https://doi.org/10.1103/PhysRevLett.105.240401},
}

@article{Candelas1980,
  author  = {Candelas, P.},
  title   = {Vacuum Polarization in {Schwarzschild} Spacetime},
  journal = {Phys. Rev. D.},
  volume  = {21},
  pages   = {2185--2202},
  year    = {1980},
  doi     = {10.1103/PhysRevD.21.2185}
}

@article{DalibardCastinMolmer1992,
  author  = {Dalibard, Jean and Castin, Yvan and M{\o}lmer, Klaus},
  title   = {Wave-Function Approach to Dissipative Processes in Quantum Optics},
  journal = {Phys. Rev. Lett.},
  volume  = {68},
  pages   = {580--583},
  year    = {1992},
  doi     = {10.1103/PhysRevLett.68.580}
}

@article{SongYaoWang2019,
  author  = {Song, Fei and Yao, Shunyu and Wang, Zhong},
  title   = {Non-Hermitian Skin Effect and Chiral Damping in Open Quantum Systems},
  journal = {Phys. Rev. Lett.},
  volume  = {123},
  pages   = {170401},
  year    = {2019},
  doi     = {10.1103/PhysRevLett.123.170401},

}

@article{HagaEtAl2021,
  author  = {Haga, Taiki and Nakagawa, Masaya and Hamazaki, Ryusuke and Ueda, Masahito},
  title   = {Liouvillian Skin Effect: Slowing Down of Relaxation Processes without Gap Closing},
  journal = {Phys. Rev. Lett.},
  volume  = {127},
  pages   = {070402},
  year    = {2021},
  doi     = {10.1103/PhysRevLett.127.070402}
}

@article{YangJiangBergholtz2022,
  author  = {Yang, Fan and Jiang, Qing-Dong and Bergholtz, Emil J.},
  title   = {Liouvillian Skin Effect in an Exactly Solvable Model},
  journal = {Phys. Rev. Research},
  volume  = {4},
  pages   = {023160},
  year    = {2022},
  doi     = {10.1103/PhysRevResearch.4.023160}
}

@article{YangZelenayovaMoligniniBergholtz2025,
  author={Yang, Fan and Zelenayova, Maria and Molignini, Paolo and Bergholtz, Emil J.},
  title={Quantum Dynamical Signatures of Non-Hermitian Boundary Modes},
  year = {2025},
  journal = {arXiv preprint arXiv:2506.16308},
  doi           = {10.48550/arXiv.2506.16308}
}

@article{GoriniKossakowskiSudarshan1976,
  author  = {Gorini, Vittorio and Kossakowski, Andrzej and Sudarshan, E. C. G.},
  title   = {Completely Positive Dynamical Semigroups of ${N}$-Level Systems},
  journal = {J. Math. Phys.},
  volume  = {17},
  number  = {5},
  pages   = {821--825},
  year    = {1976},
  doi     = {10.1063/1.522979}
}

@article{Lindblad1976,
  author  = {Lindblad, G{\"o}ran},
  title   = {On the Generators of Quantum Dynamical Semigroups},
  journal = {Commun. Math. Phys.},
  volume  = {48},
  number  = {2},
  pages   = {119--130},
  year    = {1976},
  doi     = {10.1007/BF01608499}
}

@article{Prosen2008ThirdQuantization,
  author        = {Prosen, Toma{\v z}},
  title         = {Third Quantization: A General Method to Solve Master Equations for Quadratic Open Fermi Systems},
  journal       = {New J. Phys.},
  volume        = {10},
  number        = {4},
  pages         = {043026},
  year          = {2008},
  doi           = {10.1088/1367-2630/10/4/043026},
}

@article{SinhaRavalHu2003,
  author  = {Sinha, Sukanya and Raval, Alpan and Hu, B. L.},
  title   = {Black Hole Fluctuations and Backreaction in Stochastic Gravity},
  journal = {Found. Phys.},
  volume  = {33},
  number  = {1},
  pages   = {37--64},
  year    = {2003},
  doi     = {10.1023/A:1022815724856},

}

@article{YuZhang2008,
  author  = {Yu, Hongwei and Zhang, Jialin},
  title   = {Understanding {Hawking} Radiation in the Framework of Open Quantum Systems},
  journal = {Phys. Rev. D},
  volume  = {77},
  pages   = {024031},
  year    = {2008},
  doi     = {10.1103/PhysRevD.77.024031},
}

@article{KaplanekBurgess2021,
  author  = {Kaplanek, Greg and Burgess, C. P.},
  title   = {Qubits on the Horizon: Decoherence and Thermalization near Black Holes},
  journal = {J. High Energy Phys.},
  volume  = {2021},
  number  = {1},
  pages   = {098},
  year    = {2021},
  doi     = {10.1007/JHEP01(2021)098},
}

@article{WusterSavage2007,
  author  = {W{\"u}ster, S. and Savage, C. M.},
  title   = {Limits to the Analogue {Hawking} Temperature in a {Bose}--{Einstein} Condensate},
  journal = {Phys. Rev. A},
  volume  = {76},
  pages   = {013608},
  year    = {2007},
  doi     = {10.1103/PhysRevA.76.013608},

}

@article{LombardoTuriaci2012,
  author  = {Lombardo, Fernando C. and Turiaci, Gustavo J.},
  title   = {Decoherence and Loss of Entanglement in Acoustic Black Holes},
  journal = {Phys. Rev. Lett.},
  volume  = {108},
  pages   = {261301},
  year    = {2012},
  doi     = {10.1103/PhysRevLett.108.261301},

}

@article{LombardoTuriaci2013,
  author  = {Lombardo, Fernando C. and Turiaci, Gustavo J.},
  title   = {Dynamics of an Acoustic Black Hole as an Open Quantum System},
  journal = {Phys. Rev. D},
  volume  = {87},
  pages   = {084028},
  year    = {2013},
  doi     = {10.1103/PhysRevD.87.084028},

}

@article{Franke1976,
  author  = {Franke, V. A.},
  title   = {On the General Form of the Dynamical Transformation of Density Matrices},
  journal = {Theor. Math. Phys.},
  volume  = {27},
  number  = {2},
  pages   = {406--413},
  year    = {1976},
  doi     = {10.1007/BF01051230}
}

@article{ZhangBarthel2022Criticality,
  author        = {Zhang, Yikang and Barthel, Thomas},
  title         = {Criticality and Phase Classification for Quadratic
                   Open Quantum Many-Body Systems},
  journal       = {Phys. Rev. Lett.},
  volume        = {129},
  pages         = {120401},
  year          = {2022},
  doi           = {10.1103/PhysRevLett.129.120401},

}

@article{HorstmannCiracGiedke2013,
  author        = {Horstmann, Birger and Cirac, J. Ignacio and Giedke, G{\'e}za},
  title         = {Noise-Driven Dynamics and Phase Transitions in Fermionic Systems},
  journal       = {Phys. Rev. A},
  volume        = {87},
  number        = {1},
  pages         = {012108},
  year          = {2013},
  doi           = {10.1103/PhysRevA.87.012108},

}

@article{Bravyi2005,
  author        = {Bravyi, Sergey},
  title         = {Lagrangian Representation for Fermionic Linear Optics},
  journal       = {Quantum Inf. and Comput.},
  volume        = {5},
  number        = {3},
  pages         = {216--238},
  year          = {2005},
  doi           = {10.26421/QIC5.3-3},
  eprint        = {quant-ph/0404180},
  archivePrefix = {arXiv},
  primaryClass  = {quant-ph}
}

@article{Higham1986,
  author  = {Higham, Nicholas J.},
  title   = {Computing the Polar Decomposition---with Applications},
  journal = {SIAM J. Sci. Stat. Comput.},
  volume  = {7},
  number  = {4},
  pages   = {1160--1174},
  year    = {1986},
  doi     = {10.1137/0907079}
}

@article{ShapourianShiozakiRyu2017,
  author  = {Shapourian, Hassan and Shiozaki, Ken and Ryu, Shinsei},
  title   = {Partial Time-Reversal Transformation and Entanglement
             Negativity in Fermionic Systems},
  journal = {Phys. Rev. B},
  volume  = {95},
  pages   = {165101},
  year    = {2017},
  doi     = {10.1103/PhysRevB.95.165101}
}

@article{ShapourianRyu2019,
  author  = {Shapourian, Hassan and Ryu, Shinsei},
  title   = {Entanglement Negativity of Fermions: Monotonicity,
             Separability Criterion, and Classification of Few-Mode States},
  journal = {Phys. Rev. A},
  volume  = {99},
  pages   = {022310},
  year    = {2019},
  doi     = {10.1103/PhysRevA.99.022310}
}

@article{MiyakeEtAl2013,
  author  = {Miyake, Hirokazu and Siviloglou, Georgios A. and Kennedy,
             Colin J. and Burton, William Cody and Ketterle, Wolfgang},
  title   = {Realizing the Harper Hamiltonian with Laser-Assisted
             Tunneling in Optical Lattices},
  journal = {Phys. Rev. Lett.},
  volume  = {111},
  pages   = {185302},
  year    = {2013},
  doi     = {10.1103/PhysRevLett.111.185302}
}

@article{GouEtAl2020,
  author  = {Gou, Wei and Chen, Tao and Xie, Dizhou and Xiao, Teng
             and Deng, Tian-Shu and Gadway, Bryce and Yi, Wei and Yan, Bo},
  title   = {Tunable Nonreciprocal Quantum Transport through a
             Dissipative Aharonov--Bohm Ring in Ultracold Atoms},
  journal = {Phys. Rev. Lett.},
  volume  = {124},
  pages   = {070402},
  year    = {2020},
  doi     = {10.1103/PhysRevLett.124.070402}
}

@article{LiangEtAl2022,
  author  = {Liang, Qian and Xie, Dizhou and Dong, Zhaoli and Li,
             Haowei and Li, Hang and Gadway, Bryce and Yi, Wei and Yan, Bo},
  title   = {Dynamic Signatures of Non-Hermitian Skin Effect and
             Topology in Ultracold Atoms},
  journal = {Phys. Rev. Lett.},
  volume  = {129},
  pages   = {070401},
  year    = {2022},
  doi     = {10.1103/PhysRevLett.129.070401}
}

@article{RenEtAl2022,
  author  = {Ren, Zejian and Liu, Dong and Zhao, Entong and He,
             Chengdong and Pak, Ka Kwan and Li, Jensen and Jo, Gyu-Boong},
  title   = {Chiral Control of Quantum States in Non-Hermitian
             Spin--Orbit-Coupled Fermions},
  journal = {Nat. Phys.},
  volume  = {18},
  pages   = {385--389},
  year    = {2022},
  doi     = {10.1038/s41567-021-01491-x}
}

@article{ZhaoEtAl2025,
  author  = {Zhao, Entong and Wang, Zhiyuan and He, Chengdong and
             Poon, Ting Fung Jeffrey and Pak, Ka Kwan and Liu, Yu-Jun
             and Ren, Peng and Liu, Xiong-Jun and Jo, Gyu-Boong},
  title   = {Two-Dimensional Non-Hermitian Skin Effect in an
             Ultracold Fermi Gas},
  journal = {Nature},
  volume  = {637},
  pages   = {565--573},
  year    = {2025},
  doi     = {10.1038/s41586-024-08347-3}
}

@article{WeidemannEtAl2020,
  author  = {Weidemann, Sebastian and Kremer, Mark and Helbig, Tobias
             and Hofmann, Tobias and Stegmaier, Alexander and Greiter,
             Martin and Thomale, Ronny and Szameit, Alexander},
  title   = {Topological Funneling of Light},
  journal = {Science},
  volume  = {368},
  pages   = {311--314},
  year    = {2020},
  doi     = {10.1126/science.aaz8727}
}

@article{ShiEtAl2023,
  author  = {Shi, Yun-Hao and Yang, Run-Qiu and Xiang, Zhongcheng and
             Ge, Zi-Yong and Li, Hao and Wang, Yong-Yi and Huang,
             Kaixuan and Tian, Ye and Song, Xiaohui and Zheng, Dongning
             and Xu, Kai and Cai, Rong-Gen and Fan, Heng},
  title   = {Quantum Simulation of Hawking Radiation and Curved
             Spacetime with a Superconducting On-Chip Black Hole},
  journal = {Nat. Commun.},
  volume  = {14},
  pages   = {3263},
  year    = {2023},
  doi     = {10.1038/s41467-023-39064-6}
}

@article{LappEtAl2019,
  author  = {Lapp, Samantha and Ang'ong'a, Jackson and An, Fangzhao Alex
             and Gadway, Bryce},
  title   = {Engineering tunable local loss in a synthetic lattice of
             momentum states},
  journal = {New J. Phys.},
  volume  = {21},
  number  = {4},
  pages   = {045006},
  year    = {2019},
  doi     = {10.1088/1367-2630/ab1147}
}

@article{ChengEtAl2023,
  author  = {Cheng, Dali and Lustig, Eran and Wang, Kai and Fan, Shanhui},
  title   = {Multi-dimensional band structure spectroscopy in the
             synthetic frequency dimension},
  journal = {Light Sci. Appl.},
  volume  = {12},
  pages   = {158},
  year    = {2023},
  doi     = {10.1038/s41377-023-01196-1}
}

@article{GeraceCarusotto2012,
  author  = {Gerace, Dario and Carusotto, Iacopo},
  title   = {Analog Hawking radiation from an acoustic black hole in a flowing polariton superfluid},
  journal = {Phys. Rev. B},
  volume  = {86},
  number  = {14},
  pages   = {144505},
  year    = {2012},
  doi     = {10.1103/PhysRevB.86.144505}
}

@article{NguyenEtAl2015,
  author  = {Nguyen, H. S. and Gerace, D. and Carusotto, I. and Sanvitto, D. and Galopin, E. and Lema{\^i}tre, A. and Sagnes, I. and Bloch, J. and Amo, A.},
  title   = {Acoustic Black Hole in a Stationary Hydrodynamic Flow of Microcavity Polaritons},
  journal = {Phys. Rev. Lett.},
  volume  = {114},
  number  = {3},
  pages   = {036402},
  year    = {2015},
  doi     = {10.1103/PhysRevLett.114.036402}
}
\end{document}